\documentclass{article}
\usepackage{ijcai23}

\usepackage{times}
\usepackage{soul}
\usepackage{url}
\usepackage[hidelinks]{hyperref}
\usepackage[utf8]{inputenc}
\usepackage[small]{caption}
\usepackage{graphicx}
\usepackage{amsmath}
\usepackage{amsthm}
\usepackage{booktabs}
\usepackage{algorithm}
\usepackage{algorithmic}
\usepackage[switch]{lineno}

\usepackage{amsfonts,bm,subfig,enumitem}
\newcommand{\R}{\mathbb{R}}

\title{Automating Rigid Origami Design}

\author{
Jeremia Geiger\footnote{Authors in alphabetical order}$^1$
\and
Karolis Martinkus\footnotemark[1]$^1$\and
Oliver Richter\footnotemark[1]$^{1}$\And
Roger Wattenhofer$^1$
\affiliations
$^1$ETH Zurich
\emails
jeremia.s.r.geiger@gmail.com,
\{martinkus, richtero, wattenhofer\}@ethz.ch
}

\begin{document}

\maketitle

\begin{abstract}
    Rigid origami has shown potential in large diversity of practical applications. However, current rigid origami crease pattern design mostly relies on known tessellations. This strongly limits the diversity and novelty of patterns that can be created. In this work, we build upon the recently developed principle of three units method to formulate rigid origami design as a discrete optimization problem, the rigid origami game.
    Our implementation allows for a simple definition of diverse objectives and thereby expands the potential of rigid origami further to optimized, application-specific crease patterns. We showcase the flexibility of our formulation through use of a diverse set of search methods in several illustrative case studies. We are not only able to construct various patterns that approximate given target shapes, but to also specify abstract, function-based rewards which result in novel, foldable and functional designs for everyday objects.

\end{abstract}

\begin{figure*}
\centering
  \includegraphics[width=0.95\textwidth]{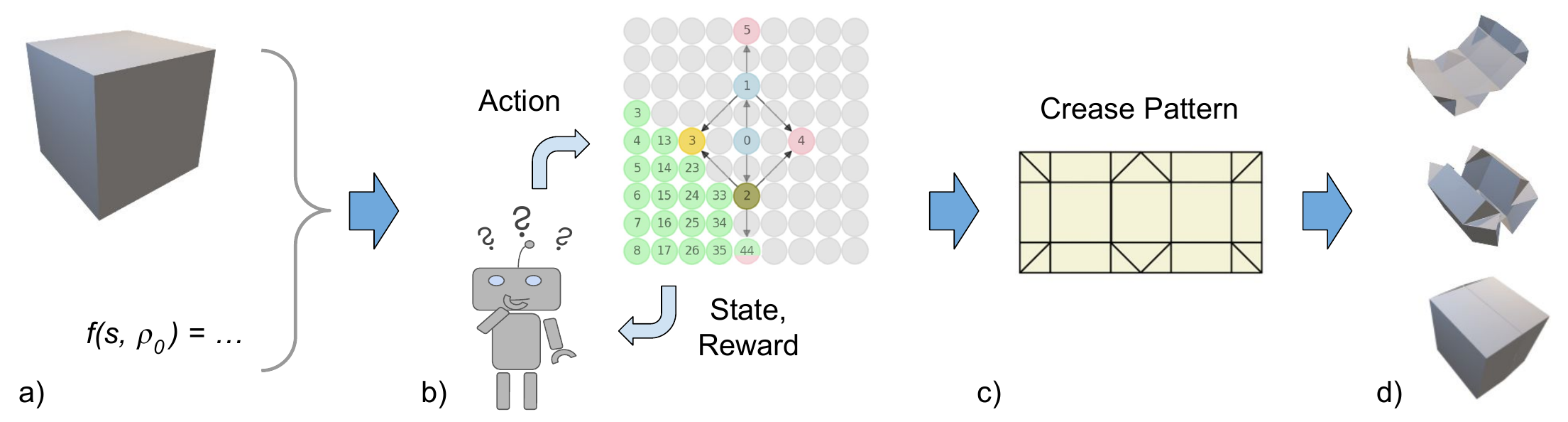}
  \caption{Overview of our approach. Given a target shape or more generally a target objective function such as desired volume (a), we formulate the inverse origami problem as a token place game on a checkerboard (b). An agent can iteratively interact with this game formulation and improve its policy with respect to the objective. The best rollout can be converted into a crease pattern (c), which can be used to rigidly fold a flat sheet into the corresponding shape that maximizes the objective (d).}
  \label{fig:teaser}
\end{figure*}

\section{Introduction}
\let\thefootnote\relax\footnotetext{Code is available at \url{https://github.com/belalugaX/rigid-origami}}
Origami may be ancient to the arts, but it is young to science.
Only in recent years, researchers began to rigorously investigate the underlying principles of folding. This has resulted in applications of origami in various fields of robotics, architecture, biomedical engineering, deployable structures, metamaterials, aerospace applications, and more \cite{engOrigamiReview2021,wang2019origami,Callens2018,Turner2015,KURIBAYASHI2006131,morgan2016approach}.

Rigidly foldable patterns, known as rigid origami, are of particular interest in practical applications since the material does not deform while folding such patterns. Even more importantly, in principle the folding motion of a rigid pattern can be induced by a single folding activation, also referred to as the patterns' degree of freedom (DOF) \cite{He2019,Luca2019-n}.
This allows for creation of shapes that are not only beautiful, but are also functional.

Traditional origami tackles the question of how a given crease pattern, defined by a set of creases drawn on a flat piece of paper, will fold. However, a more useful question is how to define a pattern such that it folds into a given target shape. This is commonly known as the inverse origami problem. Note that this is a complex problem since determining the foldability of a general crease pattern alone is NP-hard \cite{demaine2010}. Recent works started to address the inverse origami problem \cite{Demaine2017,Dudte2016,He2019}. Unfortunately, these previous works either focus on non-rigid origami and allow the pattern to have many degrees of freedom \cite{Demaine2017} or approximate shapes using only known tessellations and their variations \cite{Dudte2016,He2019,Tachi2010}. Even though it is possible to approximate some shapes very well using rigid origami tessellations, beyond the space of the tessellations there is a much larger space of general foldable patterns. Furthermore, none of these works explore the possibility of specifying not the target shape itself, but a flexible abstract objective or function that the folded shape should fulfill, which allows for even more creative freedom.


In this work, we formulate rigid origami design as a discrete optimization problem
, which allows us to search for general rigidly foldable patterns which approximate a given shape or minimize some other user-specified objective function \cite{mcadams2014novel} using a diverse set of search methods.
We treat the crease pattern as a graph where edges represent the crease lines and vertices their intersections \cite{he2019rigid}. We place vertices one after the other on a pre-defined grid to construct these graphs, see Figure~\ref{fig:teaser}.
Our formulation brings the optimization problem close to combinatorial problems, such as board games, where Reinforcement learning (RL) and other techniques have demonstrated great success  \cite{AlphaGo2016,alphazero2018,Vinyals2019}. We first derive the methodology characterizing the game environment and then expose it to various types of artificial agents for validation. Our results highlight the flexibility of our formulation and open up the possibility for future research -- both in rigid origami design and optimization methods for our application.

\section{Related Work}
In its most primitive form, a crease pattern consists of a single, center vertex and multiple leaf vertices connected to the center vertex.
The fundamental origami theorems of \citeauthor{Maekawa1983}~\shortcite{Maekawa1983} and \citeauthor{kawasaki1991relation}~\shortcite{kawasaki1991relation} determine flat foldability of such a primitive pattern, based on two characteristics of the edges: their direction of folding (the mountain-valley assignment), and their spanned planar angles (the sector angles). The foldability of multi-vertex patterns, however, has been proven to be an NP-hard problem \cite{demaine2010}, which means that there exists no efficient algorithm guaranteed to find a solution. Particularly interesting are non-developable surfaces (curved shapes), which are those that cannot be unfolded into a plane sheet.

Nevertheless, with the Origamizer, \citeauthor{Demaine2017}~\shortcite{Demaine2017} propose the state-of-the-art algorithm for solving the inverse origami problem for arbitrary target shapes. In particular, solving the inverse problem means being able to generate a crease pattern, which when folded, approximates the surface of the target up to a fixed precision.
As pointed out by \citeauthor{He2019}~\shortcite{He2019}, the Origamizer has remarkable abilities to approximate complex surfaces, but also limitations regarding our scope of rigid origami: the generated patterns have many DOF, require a sequential folding process and favor tugged panels when approximating non-developable surfaces. This makes the Origamizer solutions impractical for many engineering problems.

It has not yet been fully understood how to design rigidly foldable patterns with limited DOF \cite{He2019,Turner2015}. Thus, other state-of-the-art methods approach shape approximation from a different paradigm: It is a common practice in origami to modify a few well-known tessellation patterns, such as Miura Ori, Waterbomb-, Yoshimura-, and Resch's pattern \cite{miura1969proposition,hanna2014waterbomb,yoshimura1955mechanism,resch1970design}. Their properties are summarized and described by \citeauthor{Tachi2010}~\shortcite{Tachi2010}. In comparison to the Origamizer, methods based on known tessellations \cite{Tachi2010,He2019,Callens2018,Dudte2016} have an inherent rigid foldability and limited DOF. 
However, the main limitation of these methods is that they rely on the modified tessellations, strongly limiting the potential design space of patterns. This leads to some authors restricting themselves to approximating particular kinds of surfaces such as cylindrical ones \cite{Dudte2016}. As pointed out by \cite{Turner2015,engOrigamiReview2021}, the future prospect of origami for engineering applications relies on methods for the design of new and original rigid foldable patterns with properties specific to their field of application. 

A general design method for rigidly foldable patterns has not yet been proposed \cite{engOrigamiReview2021,Turner2015}. Nevertheless, the recently proposed principle of three units (PTU) method \cite{Luca2019-n,Luca2019-4} allows for the rule-based generation and efficient simulation of rigidly foldable patterns. The PTU implies a set of rules, such that a pattern complying with those rules is guaranteed to fold rigidly. Although the PTU does have limitations, in particular a restriction to acyclic origami graphs, it still enables the rule-based generation and efficient folding simulation of rigidly foldable patterns in large configuration spaces. 

We seek to formulate the graph design problem in a reinforcement learning (RL) setting to leverage the spectacular success of RL methods on various problems with large search spaces \cite{AlphaGo2016,alphazero2018,Vinyals2019}. RL has also been applied to various design tasks, such as designing molecules \cite{simm2020rl,simm2020symmetry} or designing computer chips \cite{mirhoseini2020chip}. However, to the bests of our knowledge, there have been no attempts so far to apply RL to rigid origami pattern discovery and the inverse origami problem in general.

Our approach also allows for a more interesting direction than simply approximating target shapes: discovering origami patterns with desired properties, such as a given surface area and height of the folded pattern. In a very nascent example of this, a specifically formulated genetic algorithm has been used to discover a simple pattern that folds a square sheet into a smaller flat sheet that has a predetermined area using a limited number of crease lines \cite{mcadams2014novel}. We explore this in a much more general form. Where our purposed method can tackle a wide array of objectives and thus applications.

\section{Rigid Origami as Discrete Optimization}
We formulate the task of designing a rigid origami crease pattern as a game in which an agent iteratively constructs the crease pattern by taking actions and receiving observations and rewards in response. Such an abstract formulation admits not only the use of reinforcement learning and evolutionary approaches but also constraint satisfaction search methods that explore the game tree.

\subsection{Notation}

We denote a crease pattern's state at time step $t$ by $s_t\in \mathcal{S}$, where $\mathcal{S}$ denotes the space of developable crease patterns which can be unfolded into a plane sheet. Every state $s\in\mathcal{S}$ can further be represented by the corresponding crease pattern graph $s=G(V,E)$, where $V$ is the set of vertices where crease lines meet and $E$ is the set of directed edges along the crease lines.

\begin{figure}[t]
    \centering
    \includegraphics[width=0.925\columnwidth]{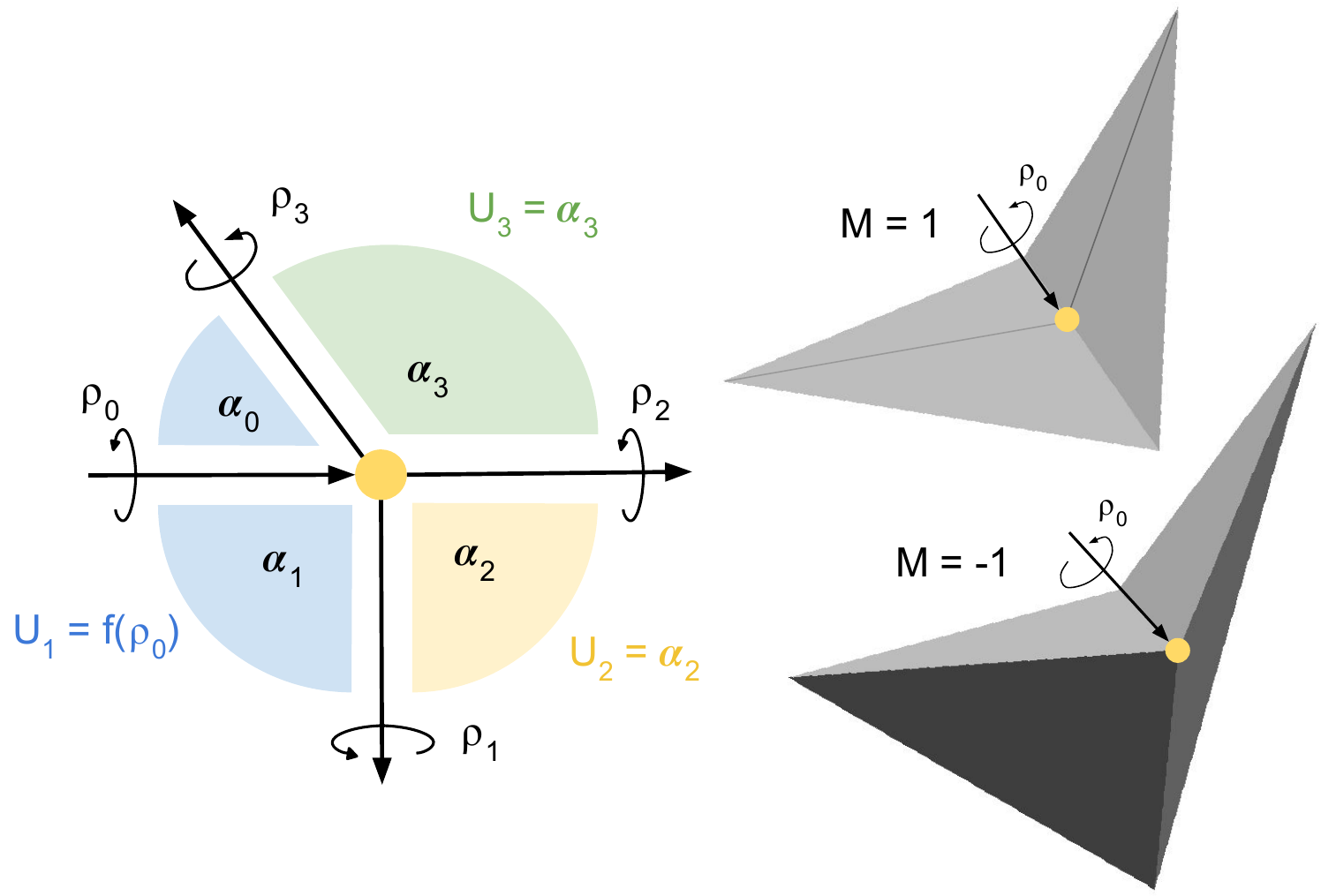}
    \caption{A degree 4 vertex with one incoming and 3 outgoing edges. Indicated are the sector angles $\mathbf{\alpha}$, the folding (dihedral) angles $\mathbf{\rho}$ and the unit angles $U_1$, $U_2$ and $U_3$. The illustrations on the right show the effect of the rigid body mode $M\in \{-1, 1\}$ for the same input angle $\rho_0$.}
    \label{fig:ptu}
\end{figure}

Every degree-$\delta$ vertex $v_i \in V$ is characterized by a set of sector angles $\boldsymbol{\alpha} = \{\alpha_i\}$ and a rigid body mode $M$, while every crease line $e_i\in E$ has a folding (dihedral) angle $\{\rho_i\}$ assigned to it. Here $\alpha_i \in (0,2\pi), \; \rho_i \in (-\pi,\pi),$ and $M \in \{-1,1\}$. Figure~\ref{fig:ptu} shows an illustration of these definitions for a simple degree 4 vertex.
Sector angles $\boldsymbol{\alpha}$ remain fixed during the folding motion (as panels are rigid), whereas the dihedral angles $\boldsymbol{\rho}$ change as the object folds according to the dynamics implied by the crease pattern. Throughout the paper, we limit our patterns to fold within a single degree of freedom (DOF), represented by the driving angle $\rho_0$.\footnote{We do allow for multiple sources in some of our starting configurations, however, all sources have the same driving angle $\rho_0$.} Note that patterns with a single driving angle are particularly appealing, as we only have to actuate one hinge to fold the whole object.
The solution of vertex kinematics for any given configuration depends on the mountain or valley assignment of the creases. This property is expressed by the rigid body mode $M$.

\subsection{PTU Kinematic Model}
The PTU \cite{Luca2019-n} provides both general rules for expanding an embedded crease pattern $s$, and an efficient kinematic model for folding simulation. 

Specifically, for a directed acyclic graph $G(V, E)$
the PTU kinematic model allows us to sequentially solve for all folding angles $\boldsymbol{\rho}$,
starting from the graph origin and finishing at the leaves. 

In its essence the PTU is based on the following observation: for a given vertex, its rigid folding motion is kinematically determinate if all but three of the adjacent edges' driving angles $\rho_i$ are known (incoming edges). The angles spanned between the three outgoing edges then define the three unit angles $U_1$, $U_2$, and $U_3$, which span a spherical triangle on the unit sphere. The PTU kinematic model now provides the formalism to solve for the outgoing edges' driving angles $\rho_j$ by means of spherical trigonometry. Hence the vertex' folding motion is fully defined.

Therefore, if we are given the driving angle $\rho_0$ and a directed acyclic graph as a crease pattern, we can sequentially calculate all unknown folding angles. As noted by \citeauthor{Luca2019-n}~(\citeyear{Luca2019-n}), this only involves forward kinematics and is thereby computationally cheaper than other methods that have to rely on matrix inversions~\cite{Tachi2010}. We refer an interested reader to~\cite{Luca2019-n} for the details of the kinematic model and focus on the graph construction here.

\subsection{The Origami Game}
\label{sec:game}
\begin{figure}[t]
    \centering
    \includegraphics[width=0.875\columnwidth]{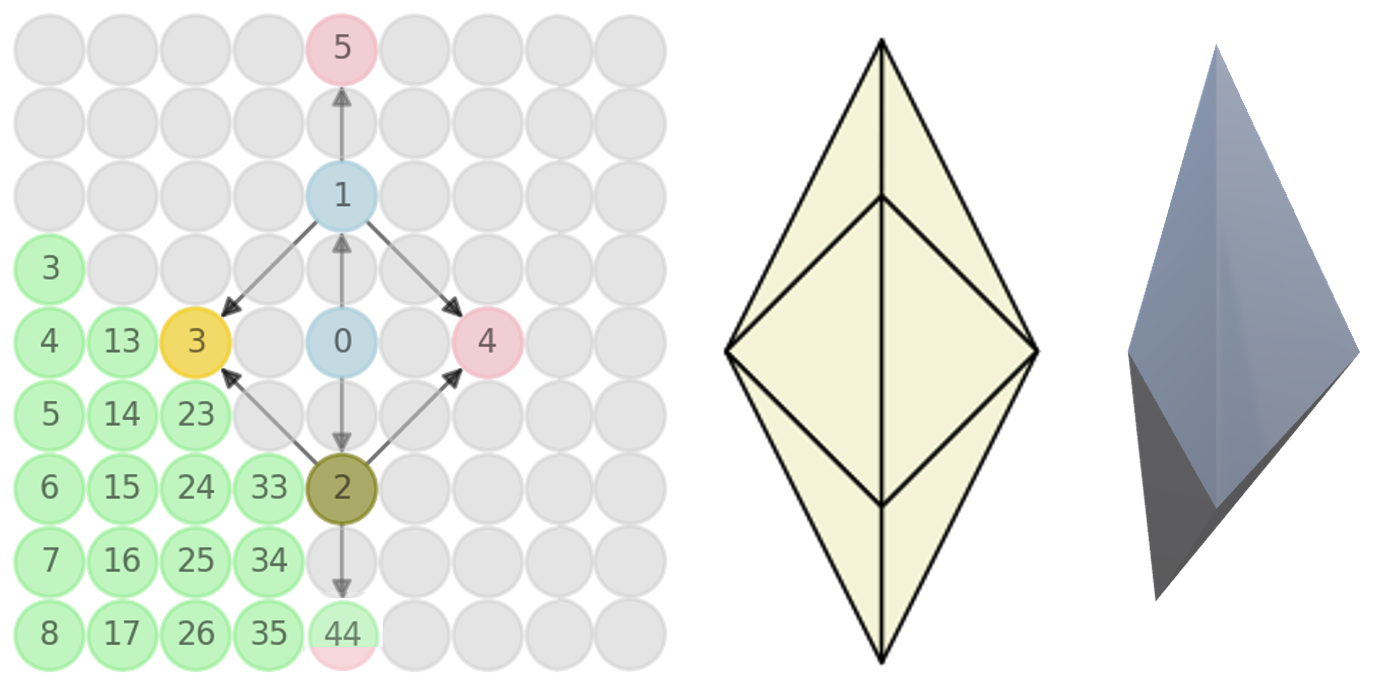}
    \caption{Origami game in a particular state (left: board, middle: creased paper, right: folded shape). The vertices on the board are color-coded as follows: Blue/Olive vertices have been extended, with the color indicating the rigid body mode $M$, pink vertices are extendable and the yellow vertex is currently selected for an extension. Green slots indicate permitted and the grey slots forbidden locations for the first outgoing edge endpoint -- these indicate the available actions. Note that node 44 is colored pink and green, as it is an extendable vertex that can be selected as an endpoint.} 
    \label{fig:game}
    \vspace{-0.55em}
\end{figure}

Given the constraints of the PTU method, we aim to formulate the task of finding a suitable crease pattern $s$ as a discrete optimization problem. More specifically, we sequentially construct a directed acyclic graph $G(V,E)$ by placing vertices $v$ and connecting crease lines $e$ on a grid board of fixed size. The board represents the unfolded paper and the discrete nature of the grid allows for efficient exploration of the infinite space of possible crease patterns.

In this formulation, we can also view the task of designing a crease pattern as playing a single-player game in which the player sequentially places vertices on the board. The board is first initialized with a simple pre-selected starting pattern, such as a line or a square, for which the size and the position can be chosen by the agent. Then, in the first phase of each turn, the player chooses a vertex that has only incoming edges so far, i.e. an extendable vertex. In the second phase, the player then chooses three locations on the board, which define the three outgoing edges of the vertex chosen before. A new vertex is created in every chosen location that did not have a vertex so far and the player is asked again to choose the next vertex to extend.
In the first phase of each turn, the player can also be given the option to choose a special \emph{source} action instead of selecting an extendable vertex. This action places a new source vertex in an empty space on the board, which receives the same external driving angle $\rho_0$ as the original source vertex, thus keeping the 1-DOF constraint. This can make for easier construction of certain shapes.

The player reaches the end of an episode in the game if either (1) a special \emph{terminate} action is played, (2) some non-foldable game state is encountered, or (3) there are no more actions available in the current game state.
See Figure~\ref{fig:game} for an illustration of the game.

\subsection{Rules and Constraints}
\label{sec:rules}
Apart from an efficient search space discretization, our board game allows us to enforce certain rules and constraints given by the PTU kinematic model and 
the rigidity of the panels being folded.

Specifically, the PTU provides a kinematic model for the folding motion of extended vertices of arbitrary degree $\delta$, given the following constraints:

\begin{enumerate}
    \item The vertex has exactly three outgoing edges and $\delta-3$ incoming edges.
    \item The triangle inequality for spherical triangles must be fulfilled throughout the entire folding motion:
    $U_{min} + U_{med} \ge U_{max}$, where $U_{min}, U_{med}, \; \textrm{and} \; U_{max}$ represent the smallest, middle and largest unit angle respectively. 
    \item The crease pattern $G(V,E)$ must be acyclic.
\end{enumerate}

\noindent Further, 
as we aim to fold rigid panels,
we can infer that:

\begin{enumerate}[resume]
    \item The crease pattern graph $G(V, E)$ must be planar, as
    crossing edges would result in overlapping panels in the flat folded starting state.
    \item Panels must not intersect with other panels during the entire folding motion.
\end{enumerate}

All constraints 1-5 are either geometrical (1,2,4,5) or topological (3) in nature. In the discrete action space, we can impose constraints 1-4 a priori through action masking, such that an agent is not eligible to take any action that would lead to a violating state.
The non-intersection of faces constraint 5, however, cannot be constrained prior to the computation of the kinematics. Hence this constraint can only be evaluated after the full extension of a vertex. 

In detail, we impose the constraints 1-5 as follows:

\begin{enumerate}
    \item By extending each vertex with exactly three child vertices. 
    \item By computing the triangle inequality for spherical triangles after the second and before the third extension of a vertex for every permitted action. We mask the actions accordingly.
    \item By masking extended vertices from the available actions.
    \item By performing an edge-edge intersection test for available actions and masking accordingly.
    \item By performing Moeller's Fast Triangle-Triangle Intersection Test \cite{Moeller1997} for a set of folding angles that mirror the folding motion. If panels intersect, we roll back to the last foldable state (the state before the failed extension) and terminate the episode.
\end{enumerate}

\subsection{Guiding the Search}
\label{sec:domain_knowledge}
Additionally to the methodological constraints outlined above which guarantee rigid foldability, our formulation also allows incorporating
additional constraints to guide the search in various aspects. This allows the user to influence and tune the result more precisely or to explore shapes with different characteristics.
Specifically:

\begin{enumerate}
    \item The board size can be chosen based on the desired complexity of the resulting shape. Smaller board sizes yield simpler solutions while larger board sizes allow for more complex shapes.
    \item Desired symmetries of the resulting shape can be incorporated by duplicating actions across symmetry axes.
    \item If a good starting pattern is known, the board can be seeded with the corresponding graph and extended from there. This can also be used to modify previously discovered or handcrafted patterns, by deleting parts of them and using the rest as a starting pattern.
    \item Finally, we can easily mask actions based on a maximal permitted crease length.
\end{enumerate}

\subsection{Objectives}
\label{sec:reward}
An important advantage of the formulation as an optimization problem is the freedom to choose the objective function. Instead of defining how the pattern should be folded, we can now focus on what we want from the resulting pattern and leave the implementation to the optimization. Specifically, given the general objective
\[\arg\max f(s,\rho_0) \]
we can design the objective function $f: \mathcal{S}\times (-\pi,\pi)\to\R$ to reflect our requirements. 

Before we give examples of explicit objective functions, we first discuss the formulation in light of the sequential construction of a solution.
That is, we seek to reward every partial solution at time step $t$ with $r_t$, such that $f(s_T,\rho_0) = \sum_{t=0}^T r_t$. Here, $T$ denotes the time step in which the episode terminates. For general objectives $f$ we can always achieve this by setting $r_T = f(s_T,\rho_0)$ and $r_{t\neq T} = 0$. However, both RL and tree search methods benefit from early feedback.
Specifically, if we wish to cut a branch in a game tree traversal at the partial solution $s_{t'}$, we can do so if 
\begin{enumerate}
    \item $r_t \leq 0 \;\forall t$, that is, any partial solution can only get worse\footnote{Note that this also implies that $f(s,\rho_0) \leq 0$ for any state $s$ and driving angle $\rho_0$. However, this can often be easily achieved by subtracting a constant that upper bounds the original objective function and using the result as an objective function.}
    \item $\sum_{t=0}^{t'} r_t < f(s^*, \rho_0)$, where $s^*$ is the best solution so far.
\end{enumerate}

Generally, we can shape rewards as $\tilde{r}_t = r_t + \Delta\Phi_t$ by adding a difference in potential $\Delta\Phi_t = \Phi(s_t, \rho_0) - \Phi(s_{t-1}, \rho_0)$ without changing the objective~\cite{reward_shaping}. However, to fulfill requirement (1) above we seek a potential function $\Phi$ that is monotonically decreasing as the game progresses, i.e. $\Phi(s_t, \rho_0) \leq \Phi(s_{t-1}, \rho_0)$. Luckily, many objectives in our setup can be naturally decomposed into a part that is monotonically decreasing with $t$ and a remainder, which can be given at the end of the episode.

To validate our approach, we first focus on shape approximation, i.e. the inverse origami problem. We then provide a range of alternative objectives to highlight the flexibility of our approach.

\subsubsection{Shape approximation}
The standard objective of the inverse origami problem is to find a crease pattern $s$ such that the folded polyhedral manifold $\mathcal{M}_{origami}\subset\R^3$ approximates a given target shape $\mathcal{M}_{target}\subset\R^3$. As a measure for the accuracy of approximation, one commonly takes the Hausdorff distance $d(X,Y)$ between sets of points $X$ and $Y$ given by 
\[d(X,Y) = \max\left\{ d_{\rightarrow}(X,Y), d_{\leftarrow}(X,Y)\right\}\]
\[d_{\rightarrow}(X,Y) = \max_{x\in X}\left(\min_{y\in Y}||x-y||_2\right)\]
\[d_{\leftarrow}(X,Y) = \max_{y\in Y}\left(\min_{x\in X}||x-y||_2\right)\]

We fix a set of sampled target points $Y\sim \mathcal{M}_{target}$ and choose $X\subset \mathcal{M}_{origami}$ as detailed later. Note that the two terms $d_{\rightarrow}(X,Y)$ and $d_{\leftarrow}(X,Y)$ model two distinct aspects: The first measures how far the folded shape is from the target shape while the second measures how well the whole target shape is covered by the folded shape. As we only add points during the construction of a crease pattern, the furthest point from the target in the folded shape can only get worse. We can therefore set $\Phi(s_t, \rho_0) = -d_{\rightarrow}(P_t, Y)$, where $P_t\in \R^{3\times|V_t|}$ denotes the spatial positions of the vertices $V_t$ in $s_t$ under driving angle $\rho_0$. We can use the vertex positions here as they reflect the corners and thereby the extremes of the folded polyhedran. The rewards are then given by $r_t=\Delta\Phi_t$ for $t\neq T$ and 
$$r_T = -\max\{d_{\rightarrow}(P_T, Y), d_{\leftarrow}(X, Y)\} - \sum_{t=0}^{T-1} r_t$$
For $r_T$ we sample $X$ from the folded shape.

\subsubsection{Abstract Rewards}

If the reward function is more abstract and does not constrain the agent to matching a given shape as close as possible the agent can create (imagine) new shapes that accomplish the goal specified by the objective function.
We sketch four kinds of handcrafted abstract objective functions to showcase the power of this approach. Specifically, starting from a flat sheet in the $xy$-plane at $z=0$, we aim to fold everyday objects:
\begin{enumerate}
    \item Bucket: To get a waterproof bucket, we maximize the smallest $z$-coordinate of the leaf nodes in the graph.
    \item Shelf: Here we maximize the area of parallel planes in the folded shape, discarding solutions with less than three parallel planes.
    \item Table: The table objective aims to get 4 points to a target $z$ value (the legs) while keeping all other points in the $z=0$ plane.
    \item Chair: A chair that is symmetrical across the $y$-axis consists of a backrest (target $z$ and $y$ value) as well as at least three legs on the ground (also with target $z$ and $y$ values).
\end{enumerate}
Note that these rewards can be augmented with certain desiderata such as a term that discourages points from sticking too far out. This augmentation can also be used for reward shaping with the same argumentation as before, points only get added, so the worst point can only get worse.
More details and the exact formulation of all objectives are provided in Appendix~\ref{appx:abstract_rewards}.

\subsection{Driving Angle Optimization}
The fitness of the final folded shape greatly depends on the driving angle $\rho_0$. Unfortunately, the optimal angle is usually not known before trying to fold the shape. In order to not constrain the agent to finding a pattern that is optimal given some particular driving angle, we instead specify a maximum driving angle $\rho_{0}^{\text{max}}$ and keep track of $10$ equally spaced driving angles from $0^{\circ}$ to  $\rho_{0}^{\text{max}}$. During each reward calculation, we calculate the reward for each of the driving angles and give the highest of those rewards to the agent. If at any point a driving angle results in an intersection, we discard it. If there are no more driving angles that do not cause an intersection the episode terminates. Note that this optimization over the driving angle is optional since we can also provide $\rho_0$ if we know it upfront from domain knowledge. However, since we intend to highlight the general applicability of our approach in our experiments we include this optimization by default.

\subsection{Search Methods}

As with the objectives, our formulation as a discrete optimization problem also allows for a diverse range of search methods. Specifically, we compare the following approaches in the shape approximation task:

\begin{itemize} 
    \item Random search (RDM)
    \item Depth-first tree search (DFTS)
    \item Breadth-first tree search (BFTS)
    \item Monte Carlo tree search (MCTS)
    \item Proximal policy optimization (PPO)
    \item Evolutionary search (EVO)
\end{itemize}

RDM simply samples actions randomly from the available actions per step.

For DFTS and BFTS, the following applies: (1) actions are ordered randomly (2) we upper bound the branching factor per node to favor a broader search and (3) both search methods follow a branch and bound paradigm, such that visited nodes are pruned if their values are less than the best episode return seen throughout the search so far. DFTS searches upwards from the leaves of the game tree. BFTS samples a fixed number of available actions per node and then traverses along the child node with the highest value.

MCTS~\cite{MCTS2008} explores the game tree through random rollouts which get biased towards regions of higher returns. In accordance with best practices, we normalize the returns to $[-1,1]$. 

PPO~\cite{PPO2017} is a model-free deep reinforcement learning algorithm, which has seen a broad adaptation in the literature as a strong baseline.
For our experiments, we rely on the PPO implementation by \citeauthor{Rllib2018}~\shortcite{Rllib2018}.

For the evolutionary algorithm (EVO) we model agents as a simple list of action values, where the list length is proportional to the board size. The first half of the list is used in the node selection phase and the other is used in the extension phase. Actions are chosen greedily with respect to the listed values of the available actions. The best-performing agents are allowed to reproduce, such that the next generation consists of randomly perturbed copies of these agents as well as some new randomly initialized agents (newcomers).

See Appendix~\ref{appx:search_details} for more details on all search methods and the hyperparameters used.

\section{Experiments}
\begin{figure}
    \centering
    \begin{minipage}{1.0\columnwidth}
        \centering
        \includegraphics[width=0.25\columnwidth]{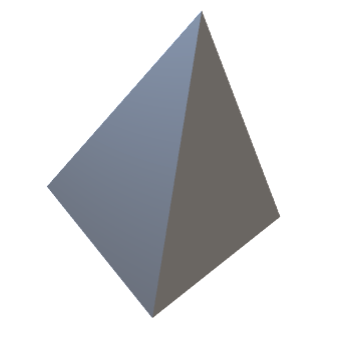}
        \includegraphics[width=0.25\columnwidth]{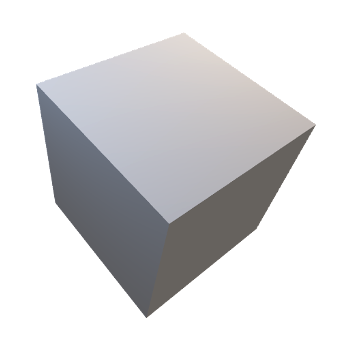}
        \includegraphics[width=0.25\columnwidth]{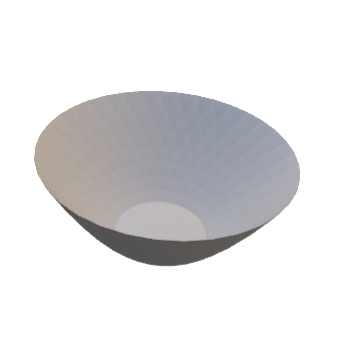}
        \includegraphics[width=0.2\columnwidth]{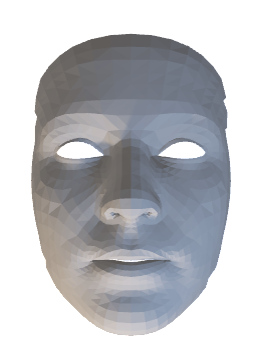}
    \end{minipage}
    \caption{Visualization of the four target shapes used: cube, pyramid, bowl, and human face.}
    \label{fig:target_shapes}
\end{figure}

First to to validate our general formulation and to benchmark different search methods we test the shape approximation task, on shapes of increasing complexity. See Figure~\ref{fig:target_shapes} for a visualization of these targets and Appendix~\ref{appx:target_shapes} for specific details. The first two shapes of cube and pyramid are developable (if cut open along the edges) and hence we expect that there exists an optimal pattern that can be modeled within the scope of our discrete environment and the PTU-based model. The bowl represents a complex, non-developable surface, that should result in more creative solutions.
Finally we push the limitations of our formulation as we try to approximate the shape of a human face, which has very complex topography. 

Secondly, we test our approach in free-form shape imagination. In this case only an abstract reward function is specified (Appendix~\ref{appx:abstract_rewards}). We aim to capture the functionality of a few standard objects, such as a bucket (maximizing volume), table (maximizing surface area at a certain height) and a shelf (maximizing collective surface area over multiple heights).

All of the experiments were performed on a server with two 10-core Intel Xeon E5-2690 v2 CPUs and 125 GB of RAM.
The environment is implemented in Open AI Gym \cite{openaigym2016}.
    
\subsection{Shape Approximation}

We ran each search method for 500 thousand interactions with the environment, always keeping track of the best pattern found so far.
To compare the search methods we evaluate the return $f(s^*, \rho_0)$ of the best pattern $s^*$ that has been found.
%
Since all our search methods are randomized algorithms, we run all searches for ten different random seeds and report the mean and the standard deviation.
The results can be seen in Table~\ref{tab:results_R}.
\begin{table}[t]
    \resizebox{\columnwidth}{!}{\begin{tabular}{l l l l l}
        \hline
            & Pyramid & Cube    & Bowl & Human Face \\
        \hline
        RDM     & $\mathbf{-0.09\pm0.00}$ & $-0.76\pm0.13$ & $-1.20\pm0.16$ & $-4.98\pm0.35$ \\
        DFTS    & $\mathbf{-0.09\pm0.00}$ & $\mathbf{-0.07\pm0.00}$ & $-0.85\pm0.08$ & $-4.38\pm0.47$  \\
        BFTS    & $\mathbf{-0.09\pm0.00}$ & $-0.07\pm0.01$ & $\mathbf{-0.75\pm0.05}$ & $-7.94\pm0.16$  \\
        MCTS    & $\mathbf{-0.09\pm0.00}$ & $-0.79\pm0.04$ & $-1.09\pm0.09$ & $-5.01\pm0.32$  \\
        PPO     & $-0.09\pm0.01$ & $-0.68\pm0.28$ & $-1.24\pm0.30$ & $-5.70\pm0.61$   \\
        EVO     & $-0.17\pm 0.25$ & $-0.44\pm0.30$ & $-1.09\pm0.25$ & $\mathbf{-3.49\pm0.61}$ \\
        \hline
    \end{tabular}}
    \caption{Summary of best episode returns $f(\rho_0,s^*)$ across shapes and search methods. The highest returns are highlighted in bold.}\label{tab:results_R}
\end{table}
Qualitatively, we visualize the folded shapes of the best patterns found in any of the runs in Figure~\ref{fig:best_approximations_per_method}. See Appendix~\ref{appx:shape_approximation_folding} for a visualization of the crease patterns and their folding motion. In the supplementary material we also provide videos of the folding motions.

Figure~\ref{fig:best_approximations_per_method} shows that all methods are able to find the best possible approximation of the pyramid, where the slight opening at the top is an artifact of the fold angle discretization.
However, for the cube, some methods struggle to find the optimal pattern. This highlights the difficulty of the optimization problem, as many crease patterns lead to well-separated local optima. 
%
Moving to the bowl we see that since the perfect approximation is now impossible, methods come up with more creative solutions. We note that DFTS and BFTS perform comparably well here, as they can leverage local optima to prune many unpromising paths.
%
Finally, the results from the human face approximations show that an evolutionary algorithm performs the best if the search space becomes excessively large. We believe this stems from the principled exploration of multiple promising solutions in the evolutionary algorithm. 
In Appendix~\ref{appx:num_env_interactions} we report the total number of environment interactions taken by each method to reach the best pattern.

In summary, we note that if the search space is moderately large, tree based search algorithms (i.e., DFTS and BFTS) perform well. However, these methods can become excessively expensive when the search space becomes larger and their initial exploration fails to make fast progress. In particular, BFTS struggles if the non-monotonic remainder of the objective is large relative to the monotonic part which is used to cut branches. Local optima are plentiful and well separated in the search space, leading policy-based methods (MCTS and PPO) to converge to a suboptimal solution prematurely. Note that the policy-based solutions generally arrive faster at their best solution, but in the end achieve worse accuracy. In contrast, the evolutionary algorithm strikes a balance between exploration and exploitation, achieving the best performance in the most complex domain of approximating a human face.

The characteristics of our search space also makes our application interesting for optimization research and we see the application of more recent approaches such as GFlowNets~\cite{GFlowNet} as a promising direction to further explore the possibilities here.

\newcommand*\rowlegendtwo[{1}]{\begin{minipage}{0.06\columnwidth}
        \rotatebox{90}{\parbox{2.2cm}{\centering #1}}
        \end{minipage}}

\begin{figure}[t]
\centering
    \rowlegendtwo{RDM}
    \begin{minipage}{0.93\columnwidth}
        \includegraphics[width=0.23\columnwidth]{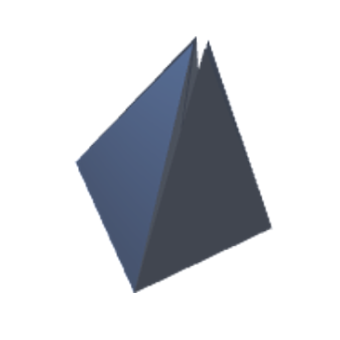}
        \includegraphics[width=0.22\columnwidth]{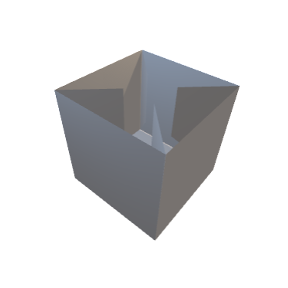}
        \hspace*{1mm}
        \includegraphics[width=0.23\columnwidth]{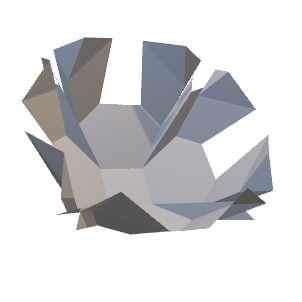}
        \includegraphics[width=0.23\columnwidth]{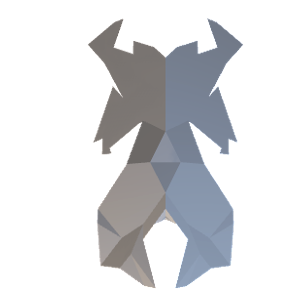}
    \end{minipage}
    \vspace*{-7mm}
    
    \rowlegendtwo{DFTS}
    \begin{minipage}{0.93\columnwidth}
        \includegraphics[width=0.23\columnwidth]{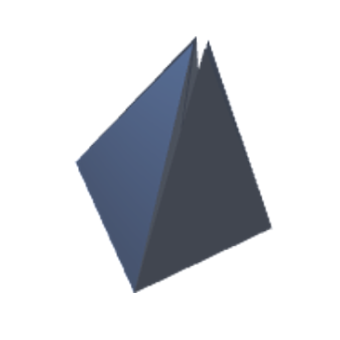}
        \includegraphics[width=0.22\columnwidth]{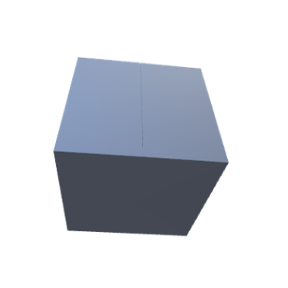}
        \hspace*{1mm}
        \includegraphics[width=0.23\columnwidth]{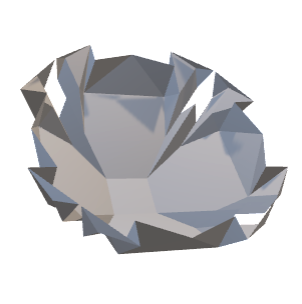}
        \includegraphics[width=0.23\columnwidth]{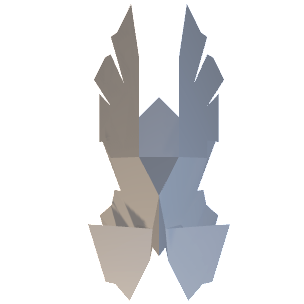}
    \end{minipage}
    \vspace*{-7mm}
    
    \rowlegendtwo{BFTS}
    \begin{minipage}{0.93\columnwidth}
        \includegraphics[width=0.23\columnwidth]{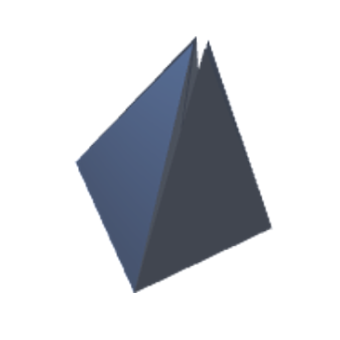}
        \includegraphics[width=0.22\columnwidth]{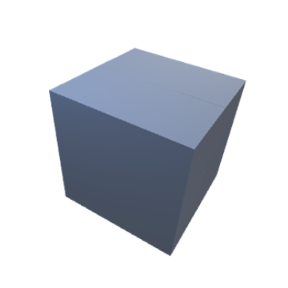}
        \hspace*{1mm}
        \includegraphics[width=0.23\columnwidth]{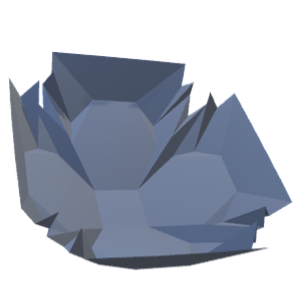}
        \includegraphics[width=0.23\columnwidth]{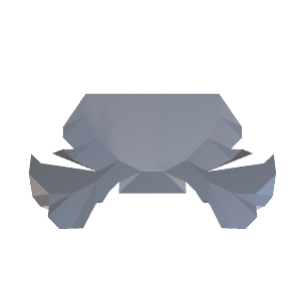}
    \end{minipage}
    \vspace*{-7mm}
    
    \rowlegendtwo{MCTS}
    \begin{minipage}{0.93\columnwidth}
        \includegraphics[width=0.23\columnwidth]{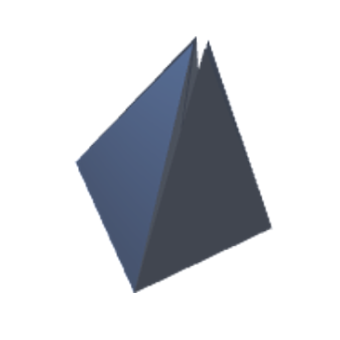}
        \includegraphics[width=0.22\columnwidth]{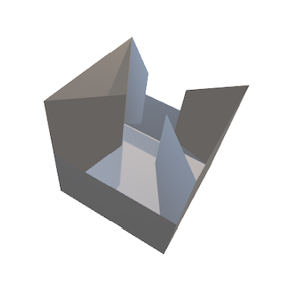}
        \hspace*{1mm}
        \includegraphics[width=0.23\columnwidth]{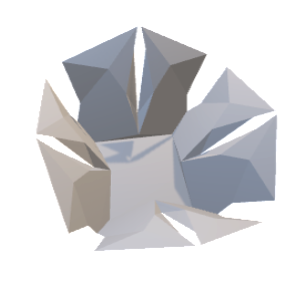}
        \includegraphics[width=0.23\columnwidth]{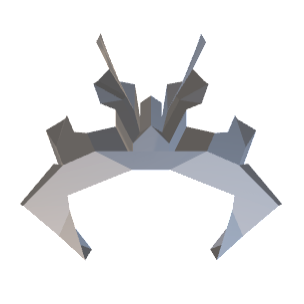}
    \end{minipage}
    \vspace*{-7mm}
    
    \rowlegendtwo{PPO}
    \begin{minipage}{0.93\columnwidth}
        \includegraphics[width=0.23\columnwidth]{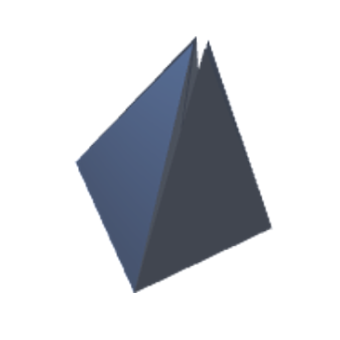}
        \includegraphics[width=0.22\columnwidth]{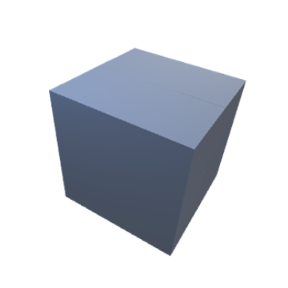}
        \hspace*{1mm}
        \includegraphics[width=0.23\columnwidth]{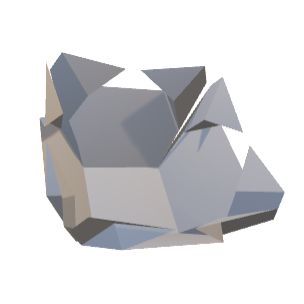}
        \includegraphics[width=0.23\columnwidth]{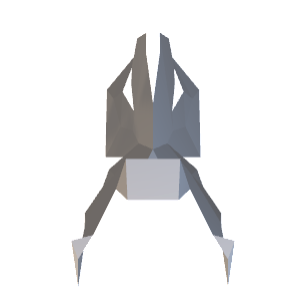}
    \end{minipage}
    \vspace*{-7mm}
    
    \rowlegendtwo{EVO}
    \begin{minipage}{0.93\columnwidth}
        \includegraphics[width=0.23\columnwidth]{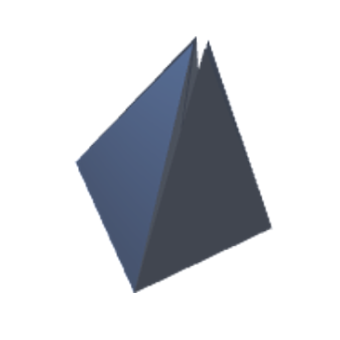}
        \includegraphics[width=0.22\columnwidth]{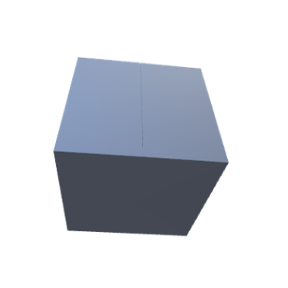}
        \hspace*{1mm}
        \includegraphics[width=0.23\columnwidth]{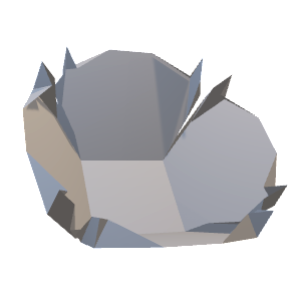}
        \includegraphics[width=0.23\columnwidth]{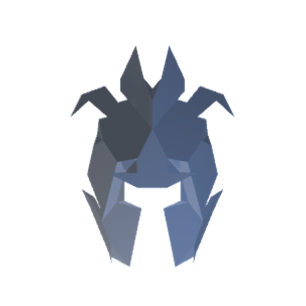}
    \end{minipage}
    
    \caption{The figure shows the best approximations found for each method over all of the ten runs, with the overall best approximations across all methods highlighted in blue.}
    \label{fig:best_approximations_per_method}
\end{figure}

\subsection{Shape Imagination}
\label{sec:imagination}

With shape imagination we address the second use case of our environment: instead of approximating a given target shape, we utilize the environment for the more abstract generative design tasks. 
Specifically, we base ourselves on the good performance of the evolutionary algorithm established before and try out the abstract objectives introduced in Section~\ref{sec:reward}. We set the board size to a moderate $13\times 13$ and run our evolutionary algorithm for 500 thousand interactions with the environment. Experiment details are given in Appendix~\ref{appx:furniture_details}.

\begin{figure}
    \centering
    \begin{minipage}{1.0\columnwidth}
        \includegraphics[width=0.24\columnwidth]{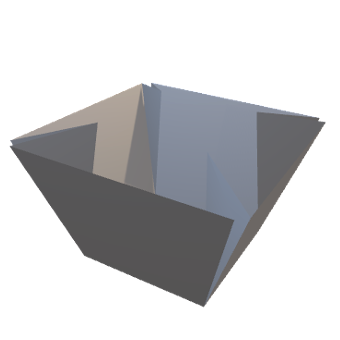}
        \includegraphics[width=0.24\columnwidth]{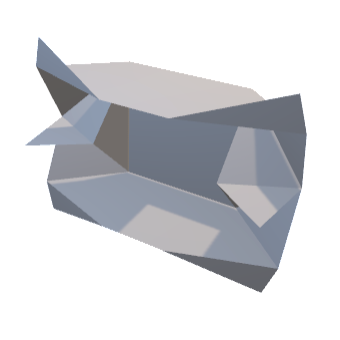}
        \includegraphics[width=0.24\columnwidth]{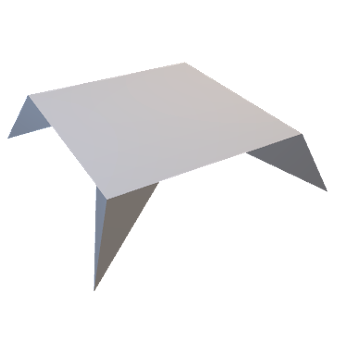}
        \includegraphics[width=0.24\columnwidth]{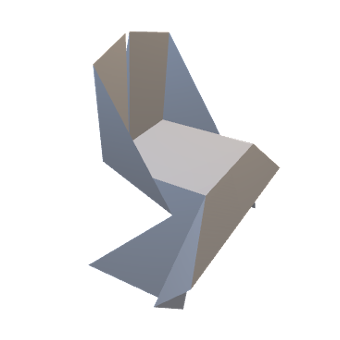}
    \end{minipage}
    \begin{minipage}{1.0\columnwidth}
        \includegraphics[width=0.24\columnwidth]{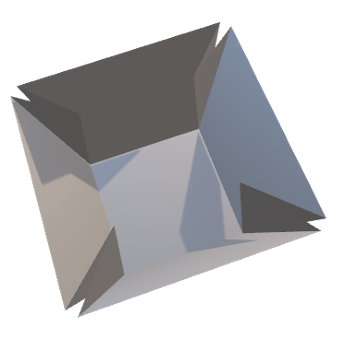}
        \includegraphics[width=0.24\columnwidth]{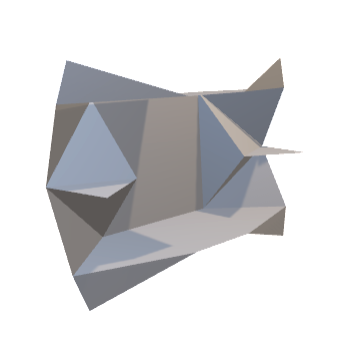}
        \includegraphics[width=0.24\columnwidth]{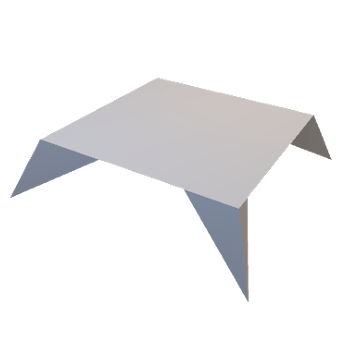}
        \includegraphics[width=0.24\columnwidth]{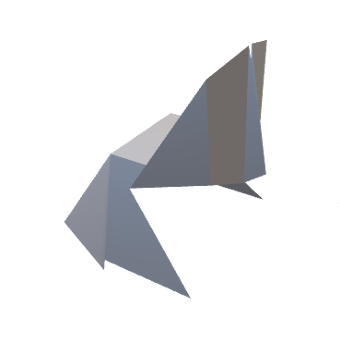}
    \end{minipage}
    \caption{Solutions found on the abstract objectives. From left to right: bucket, shelf, table, chair. The two rows display two viewing angles. 
    }
    \label{fig:abstract_objective_results}
\end{figure}

The results of the best pattern found are visualized in Figure~\ref{fig:abstract_objective_results}. See also Appendix~\ref{appx:furniture_folding} for a visualization of the crease patterns and their folding motion. These results show that our setup allows us to find interesting shapes that reflect our specifications while being developable from a flat surface using a single degree of freedom. In the context of furniture, this means that all these objects can be folded flat and stacked for storage or transportation. Moreover, the single degree of freedom makes them easy to fold, as there is a single continuous motion that leads to the folded shape. We see that there is a lot of potential in our setup for customized objectives when the design specifics are left to the optimization.

\subsection{Changing Constraints}
\label{appx:additional_imagined_shapes}

\begin{figure}[t]
    \centering
    \includegraphics[width=0.975\columnwidth]{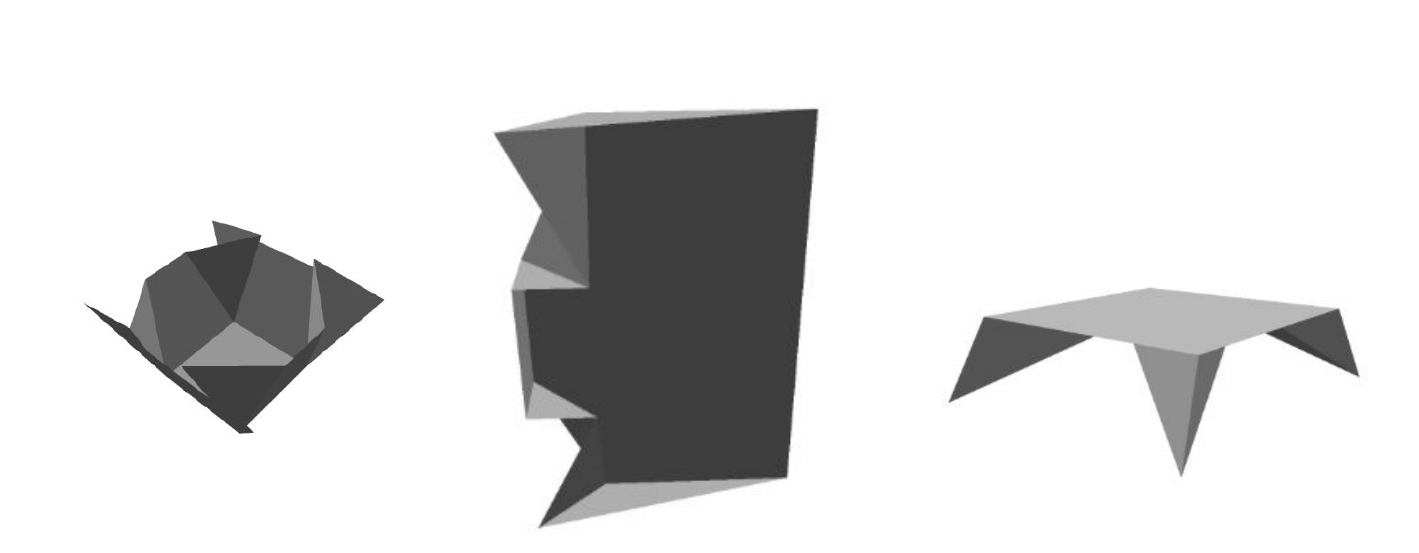}
    \caption{Additional results on the abstract objectives for different symmetry constraints. From left to right: Bucket with only $x$ and $y$ symmetry axes, shelf with only the $y$ symmetry axis, table with only $x$ and $y$ symmetry axes.}
    \label{fig:additional_imaginations}
\end{figure}

As discussed in Section~\ref{sec:domain_knowledge}, if we want to change up the results that we get, besides just changing the random seed used for the search, we can also impose different constraints on the solution space.  In the previous section we utilized stricter constraints, for example bucket and table were supposed to be perfectly symmetric, while the shelf was supposed to be symmetric in both $x$ and $y$ axes. In Figure~\ref{fig:additional_imaginations} we showcase the results that we get, if these constraints are relaxed. The other settings remain the same (Appendix~\ref{appx:furniture_details}), only for the shelf the board size is reduced to $9\times 9$ to counter the search space increase resulting from the removed symmetry.

\section{Conclusion}

In this paper, we have described an environment for the iterative search of rigidly foldable crease patterns. We discretize the search space to get an efficient model for exploration and constraint imposition. Furthermore, we have demonstrated its utility and limitations for shape approximation by evaluating various search methods, including branch and bound searches, reinforcement learning, and an evolutionary algorithm. We also highlight the potential of our environment for use with customizable objectives and use of constraints to allow the user to guide the search in a desired direction. The proposed rigid origami design environment could prove useful not only in creative endeavors of rigid origami pattern design but also for the application of novel optimization algorithms to origami design and arts.


\bibliographystyle{named}
\bibliography{ijcai23}

\clearpage
\newpage
\appendix

\section{Search Method Details}
\label{appx:search_details}
Here we provide further details of the search methods used in our experiments. All methods act within the \emph{playable area} $B$ of the board. That is, we limit the area in which the agent can act on the board according to the symmetries defined: If we consider the $y$ axis as a symmetry line, we limit the playable area to the left half of the board. If additionally to the symmetry across the $y$ axis we also enforce symmetry across the $x$ axis, we limit the playable area to the top left quarter of the board. The additional $xy$ symmetry axis used for the bowl, bucket, and table is enforced with the corresponding action masking that shrinks the playable area to $\sim 1/8$ of the original board size. Note that we keep the board positions on the symmetry axes within the playable area.

\subsection{RDM}
The random baseline simply samples in every step $t$ an action $a_t$ uniformly from the set of available actions in the state $s_t$. That is, as in all methods, we mask actions according to constraints and domain knowledge (see Sections~\ref{sec:rules} and~\ref{sec:domain_knowledge}) and sample from the remaining actions. In Figure~\ref{fig:game} we visualize an example of this action set in green.  

\subsection{DFTS}

To explore a larger part of the search space, we run 10 searches of 100'000 steps each. In each trial, we start by randomly playing a game until the end and storing all states that were visited along the way. From the final state, we back up and sample a new action in the penultimate state, traversing the game tree in a depth-first manner. Throughout the search, we keep track of the best return seen so far and immediately back up from a state if the reward trace leading to it sums to less than the best-seen return. To further encourage exploration, we limit the number of children sampled in each state to 10, backing up once this threshold is crossed.

\subsection{BFTS}

The setup for breadth-first tree search is similar to the depth-first tree search: we run 10 searches of 100'000 steps each, backing up if either the current reward trace is less than the best seen so far or the branching threshold of 10 explored children is crossed. However, instead of traversing the tree depth-first, we traverse it breadth-first. That is, in every new state we first sample up to 10 actions uniformly at random, from the available ones, evaluate the states which result from performing these actions and then move to the state which yielded the highest evaluation. If we back up to a visited state, we move to the child that scored second-highest and so on.

\subsection{MCTS}
The Monte Carlo Tree Search algorithm by~\citeauthor{MCTS2008}~\shortcite{MCTS2008} is divided into four phases: selection, expansion, simulation, and back-propagation. This algorithm has been successfully applied to game problems ~\cite{alphazero2018}. We augment the implementation provided by ~\citeauthor{Rllib2018}~\shortcite{Rllib2018} with the following default configuration settings: For each step in the actual environment, $100$ full episode simulations are performed. During the simulation phase, we traverse the tree by sampling available actions from a uniform distribution, which is distorted by an additive Dirichlet noise of $0.03*\text{Dir}(0.25)$.
For the traversal in the actual environment, we follow a greedy policy, by always selecting the most visited child. The simulation game tree is preserved throughout an episode, such that the following simulations can build on the previous findings. During the back-propagation phase, the reward is normalized as follows: $r_{norm} = \frac{2\cdot r_{t}}{r_{min}}+1$, where $r_{min}$ is the penalty received in a non-foldable state. Since the rewards in the shape approximation are upper bounded by zero, this ensures that $r_t \in [-1,1]$ for all $t$. 
Additionaly we apply the following settings: $temperature=1.5$ and a coefficient $c_{puct}=1.0$ to balance exploration and exploitation.

\subsection{PPO}

Proximal policy optimization~\cite{PPO2017} is a policy gradient-based deep reinforcement learning approach that optimizes a surrogate objective inspired by trust-region optimization \cite{schulman2015trust}. 

We use the implementation provided by~\citeauthor{Rllib2018}~\shortcite{Rllib2018}, modeling the agent as a simple multilayer perceptron with two layers of 256 neurons each. We keep most hyperparameters at the default values and only adjust some to fit our hardware. That is, we set the SGD mini-batch size to 64 and the training batch size to 2'050. We also set the entropy coefficient to 0.01 (default is 0) to encourage better exploration.

As input to the agent we provide an encoding of the state $s$ in a third order state tensor  $O_{i,j,k} = \{-1,0,1\}^{w \times h \times (wh+2)}$, where $w$ is the board width and $h$ is the board height. In more detail:
\begin{itemize}
    \item $O_{i,j,k=0}.$ The first slice of the elements encodes the rigid body mode $M\in \{-1,1\}$, where $i,j$ denotes the vertex position on the grid.
    \item $O_{i,j,k=n}.$ The last slice indicates the current vertex $i,j$, which is selected for extension in the second phase.
    \item $O_{i,j,k=1,...,k=n-1}.$ In addition to their position, vertices are enumerated along $k$ in order of their appearance. For a given vertex position $i,j$, the vector $O_{i,j,k=1,...,wh}$ encodes the vertex adjacency. The values $\{-1, 1\}$ indicate incoming, or outgoing edges respectively. 
\end{itemize} 

Similar to the other approaches we mask invalid actions from the probability distribution output over the action space.

\subsection{EVO}

In the evolutionary algorithm, we keep track of a population of 128 agents, each represented by a list of action values $\mathbf{q}\in \R^{m}$ where $m = 4d$ and $d$ is the size of the playable area. To give an example, the playable area for a $13\times 13$ board with the $x$ and $y$ axis as symmetry lines has a size of $d=7^2 = 49$. 
The size of the action value vector $m$ is $4\cdot d$ to have $2d$ action values for the node selection phase and $2d$ action values for the node extension phase, where factor 2 accounts for the selection of the rigid body mode $M$. As such, every action value corresponds to a specific action that becomes available or masked depending on the state of the game. Agents act by always choosing the action which has the highest value among the available actions.

At initialization, we sample each agents' value vector $\mathbf{q}$ from an $m$-dimensional isotropic normal distribution $\mathbf{q}\sim\mathcal{N}(\mathbf{0},\mathbf{I})$ where $\mathbf{I}$ represents the $m\times m$ identity matrix. In each iteration of the evolution, we then
\begin{enumerate}
    \item evaluate each agent in the population by playing out an episode. Here we set the evaluation of agents which yielded the same action sequence as another agent in the current generation to $-\infty$. This is done to encourage diversity. We then
    \item order agents according to their achieved evaluations
    \item take the top $25\%$ as parent population for the next generation, discarding the rest
    \item mutate the parents by adding a small random perturbation $\epsilon_1\sim \mathcal{N}(\mathbf{0},\sigma_1\mathbf{I})$
    \item copy the mutated parents twice as offspring, where each copy is subject to another additive random perturbation with $\epsilon_2\sim \mathcal{N}(\mathbf{0},\sigma_2\mathbf{I})$ and $\epsilon_3\sim \mathcal{N}(\mathbf{0},\sigma_3\mathbf{I})$ respectively.
    \item Fill up the remaining $25\%$ of the population with newcomers $\mathbf{q}\sim\epsilon_2\sim \mathcal{N}(\mathbf{0},\mathbf{I})$ and start the next iteration (go back to 1).
\end{enumerate}
We iterate until the total sum of steps taken by all agents crosses the predefined step limit of 500 thousand interactions.

The perturbation hyperparameters were set to $\sigma_1=0.1$, $\sigma_2=0.5$ and $\sigma_3=1.0$. These hyperparameters, as well as the population size of 128 agents and the threshold at $25\%$ were initial guesses that worked out well enough. We did not tune these hyperparameters in any way, therefore results might even improve for a different set of hyperparameters.

\section{Target Shapes}
\label{appx:target_shapes}
For a more formal definition of the four target shapes in Figure \ref{fig:target_shapes} refer to the list as follows:

\begin{itemize}
    \item Pyramid. The pyramid is defined by a polygonal mesh, spanned by a set of five vertices:\\
        $\{(2,2,0),(-2,2,0),(-2,-2,0),(2,-2,0), (0,0,\sqrt{8})\}$.
    \item Cube. A cube with edge length $b=2$, as expressed in units of the discrete board grid.
    \item Bowl. The curved surface of the bowl is defined by a paraboloid: 
    $z = (0.2x^2 + 0.2y^2 -0.6)$, for $\sqrt{x^2 + y^2}\in (1.6,5.0)$. The bottom of the bowl ($\sqrt{x^2 + y^2}<1.6$) is an adjacent circular disk of radius 1.6.
    \item Human Face. The polygonal mesh of a human face, slightly edited (trimmed in the depth dimension) from Apple ARKit. \footnote{Apple ARKit \url{https://developer.apple.com/augmented-reality/arkit/}}
\end{itemize}

\section{Explanation of Shape-specific Settings}
\label{appx:exp_of_shape_settings}
As our agents cannot observe the target shape, it is beneficial to indirectly tell them some important characteristics of the shape to ease the task. This is done by introducing additional constraints based on domain knowledge as discussed in Section~\ref{sec:domain_knowledge}.
The list below provides a detailed explanation of the experimental settings that are summarized in Table~\ref{tab:experiments}:

\begin{table}[t]
    \resizebox{\columnwidth}{!}{
    \begin{tabular}{l l l l l l}
        \hline
        Target  &  Board Size &  Start Pattern & Symmetry Axes & $cl_{max}$ \\
        \hline
        pyramid     &  (9,9)    & square        & x,y       & $\infty$\\
        cube        &  (9,9)    & square        & x,y       &  $\infty$\\
        bowl        &  (25,25)  & square        & x,y,xy    & 2.9\\
        human face  &  (25,25)  & single crease  & y        & 2.9\\
        \hline
    \end{tabular}
    }
    \caption{Experimental settings overview: we impose domain specific constraints on the four target shapes, regarding board size, symmetry, initial configuration and the maximal permitted length of crease lines $cl_{max}$.}
    \label{tab:experiments}
\end{table}

\begin{enumerate}
    \item The board size is chosen in relation to the size of the target, and the complexity of its surface topography.
    \item We initialize the pyramid, cube, and bowl with a square initial pattern at the center of the board. We fix the size of the square for the pyramid and the cube according to the target shape. For the bowl we let the agent choose the size of the initial square.
    \item The human face is initialized with a single crease line just above the tip of the nose.
    \item Symmetry constraints are imposed according to the symmetry of the targets. That is, for the pyramid and the cube we impose symmetry across the $x$ and $y$ axis, while for the face we only have symmetry across the $y$ axis. For the bowl, apart from the symmetry across the $x$ and $y$ axis, we add an additional $xy$-axis symmetry.
    \item We set $\rho_0^{\text{max}}=\pi$. Since the agent is also free to choose the rigid body mode $M$ of the source creases, this effectively gives it the ability to optimize over the full range $\rho_0\in (-\pi,\pi)$.
    \item For the pyramid, the maximum permitted crease length $cl_{max}$ remains unconstrained.
    \item For the cube, the maximum permitted crease length $cl_{max}$ is bound by its edge length.
    \item For the bowl and the human face we set a maximum permitted crease length such that the agent is encouraged to build more details.
    \item Additionally, we adjust the reward signal $r$ for the two closed surfaces of the pyramid and the cube as follows: any parts of the folded surface which are fully enclosed by the target surface are excluded from the reward computation. This is done as we normally do not care about the folds which lie inside the closed surface if the closed surface is our target.
\end{enumerate}

\section{Number of Environment-interactions}
\label{appx:num_env_interactions}
Number of environment-interactions after which each method finds its best pattern can be seen in 
Table \ref{tab:results_T}. PPO and MCTS tend to find their best pattern quite quickly, but sometimes only find a local minimum (see Table~\ref{tab:results_R}).
\begin{table}[t]
\centering
    \resizebox{1.0\columnwidth}{!}{\begin{tabular}{l l l l l}
        \hline
            & Pyramid & Cube    & Bowl & Human Face \\
        \hline
        RDM & $130\pm115$ & $73\pm89$ & $249\pm187$ & $193\pm93$ \\
        DFTS & $55\pm29$ &  $96\pm92$ & $371\pm99$ & $256\pm112$ \\
        BFTS & $83\pm24$ & $94\pm48$ & $380\pm62$ & $335\pm145$ \\
        MCTS & $22\pm20$ & $159\pm169$ & $233\pm136$ & $314\pm143$ \\
        PPO & $13\pm4$ & $151\pm162$ & $83\pm109$ & $297\pm237$ \\
        EVO & $59\pm68$ & $236\pm150$ & $177\pm121$ & $376\pm125$  \\
        \hline
    \end{tabular}}
    \caption{Environment-interactions (in thousands) to find the best pattern.}
    \label{tab:results_T}
\end{table}

\section{Shape Folding}
\label{appx:shape_approximation_folding}
\begin{figure*}[t]        
\centering
    \subfloat[Pyramid crease pattern]{
        \centering 
        \includegraphics[width=0.475\textwidth]{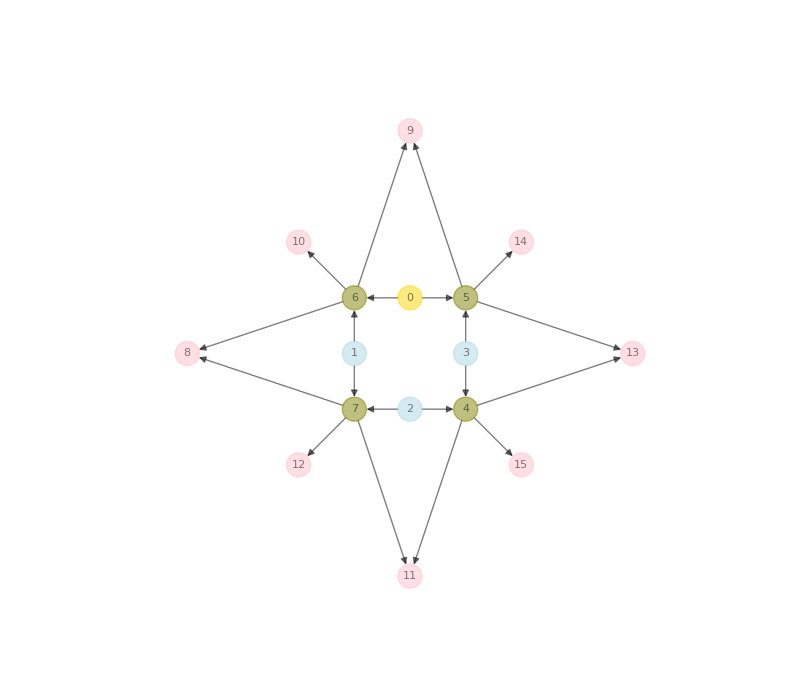}
        \label{fig:shape_approx_patterns_a}
    }
    \hfill
    \subfloat[Cube crease pattern]{
        \centering 
        \includegraphics[width=0.475\textwidth]{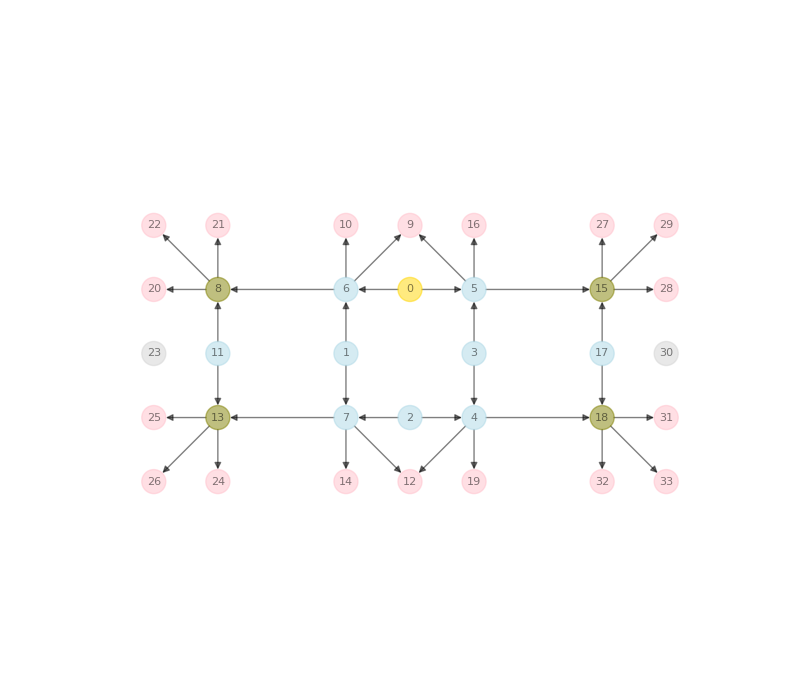}
        \label{fig:shape_approx_patterns_b}
    }
    \vskip\baselineskip
    \subfloat[Bowl crease pattern]{
        \centering 
        \includegraphics[width=0.475\textwidth]{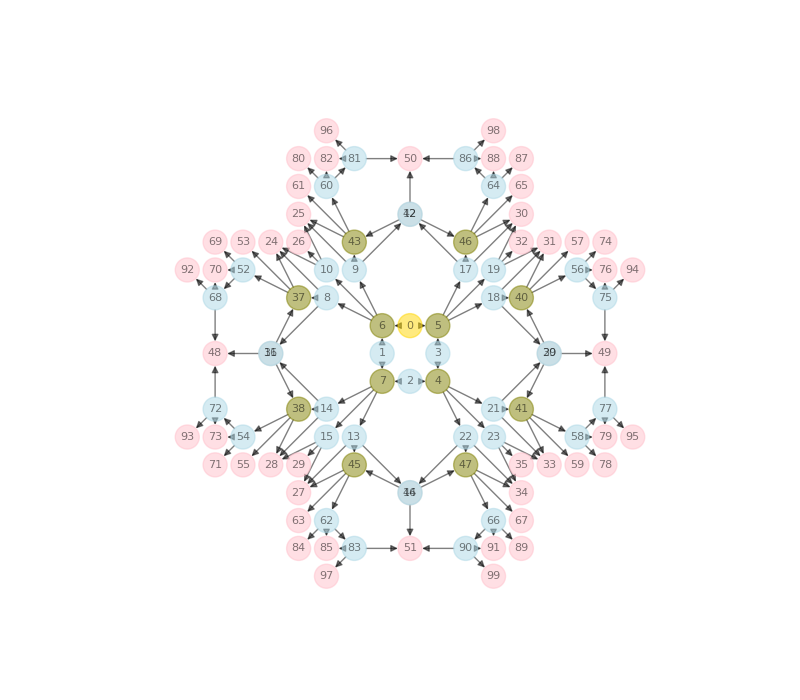}
        \label{fig:shape_approx_patterns_c}
    }
    \hfill
    \subfloat[Face crease pattern]{ 
        \centering 
        \includegraphics[width=0.475\textwidth]{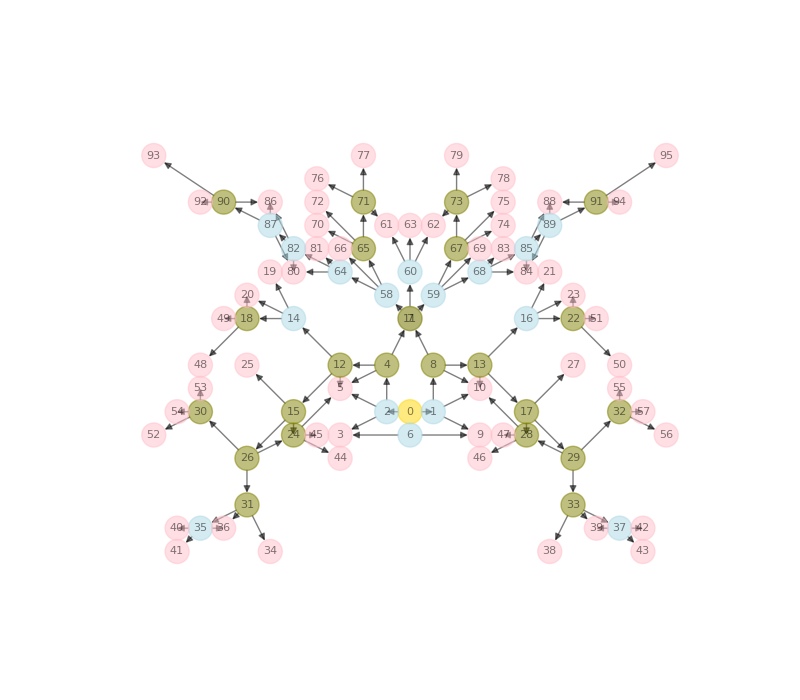}
        \label{fig:shape_approx_patterns_d}
    }
    \caption{The best approximations for the four target shapes: (a) pyramid, (b) cube, (c) bowl, and (d) face.} 
    \label{fig:shape_approx_patterns}
\end{figure*}

The best shape approximation crease patterns can be seen in Figure~\ref{fig:shape_approx_patterns} and their folding motion can be seen in Figure~\ref{fig:folding_motion}. We also provide videos of the folding motion in the supplementary material.

\section{Abstract Reward Details}
\label{appx:abstract_rewards}
Here we provide more details on the abstract reward functions used to specify the desiderata of the everyday objects. Since the evolutionary algorithm does not rely on intermediate rewards $r_t$ for $t<T$, we mainly focus on the formulation of objective $f(s_T,\rho_0)$ and simply set $r_t=0$ for $t<T$ and $r_T=f(s_T,\rho_0)$. Note however that many objectives can be decomposed and shaped if one wishes to use other search algorithms.

\subsection{Bucket}

For the waterproof bucket, we aim to maximize the $z$-value of the outermost vertices. That is, we take the vertices $v\in V_T$ which have not been extended (pink/yellow vertices in Figure~\ref{fig:game}) and use the minimum of their $z$ values in the folded shape as a reward. To encourage compact shapes we only reward buckets whose max $z$ value of the outer points is at most twice the min $z$ value and whose points have a max norm in the $xy$ plane which is at most twice the min norm in the $xy$ plane.

\subsection{Shelf}

The idea behind the shelf reward is to maximize the area where one can place things onto the shelf. To this end, we take the surface normals of the triangulated folded mesh and evaluate the pairwise cosine-similarity to see which surfaces are parallel to each other. We further calculate the distance between parallel surfaces to distinguish triangles from the same shelf level from triangles of different shelf levels. We do not reward shelves with less than three levels to exclude trivial solutions. For a shelf with at least three levels, we sum up the areas of the triangles within a plane and take the minimum of these surface areas as a reward.

\subsection{Table}

For the table we set a target $z$ value of 2.5 and calculate the mean absolute distance between the $z$ value of the 4 highest points and this target. For all other points we calculate the mean absolute difference between their $z$ value and 0, to keep them in the table plain. The final reward is the negative sum of these mean distances, i.e.
$$f(s_T, \rho_0) = -\frac{1}{4}\sum_{p\in P_T^{max}} |p_z - 2.5| - \frac{1}{|V_T| - 4}\sum_{p\in P_T\setminus P_T^{max}} |p_z|$$
where $P_T^{max}$ are the points with maximal $z$ value and $|V_T|$ is the number of vertices/points in the graph.

\subsection{Chair}
\label{appx:chair_reward}
To encourage compact solutions, we penalize solutions that have points in the folded shape with $x$ or $y$ coordinate bigger than 4, that is, any point outside an area that is 4 times as large as the starting square. We further penalize solutions that do not have their 3 lowest points at approximately the same height or have them only on one side of the board. This is to ensure that we have at least three legs that provide some sort of stability. We then categorize all points above the $z=0$ plane as backrest points and calculate their mean absolute difference to a target $y$ value of 2.1, i.e., they should lie just next to one edge of the starting square.
Further we add the absolute difference between the $z$ value of the highest point and 4 to encourage the backrest to achieve a certain height. Finally, the leg points, i.e., the 3 lowest points are also subject to corresponding target $z$ and $y$ values. Formally, we have
$$L_{legs} = \frac{1}{3}\sum_{p\in P_T^{legs}} |p_z - (-4)| + ||p_y| - 2.1| $$
$$L_{rest} = |p_z^{max} - 4| + \frac{1}{|P_T^{z>0}|}\sum_{p\in P_T^{z>0}} ||p_y| - 2.1| $$
$$f(s_T, \rho_0) = - L_{legs} - L_{rest}$$
Here $P_T^{legs}$ is the set of the 3 points with the smallest $z$ value, $p_z^{max}$ is the largest $z$ value among all of the points and $P_T^{z>0}$ is the set of points with a $z$ value larger than 0. Note the slight abuse of notation: we use $|P_T^{z>0}|$ to denote the number of points in set $P_T^{z>0}$, while the other uses of $|\cdot|$ represent the absolute value function.

\section{Shape Imagination Experiment Setup}
\label{appx:furniture_details}

\begin{figure}[h]
    \centering
    \includegraphics[width=\columnwidth]{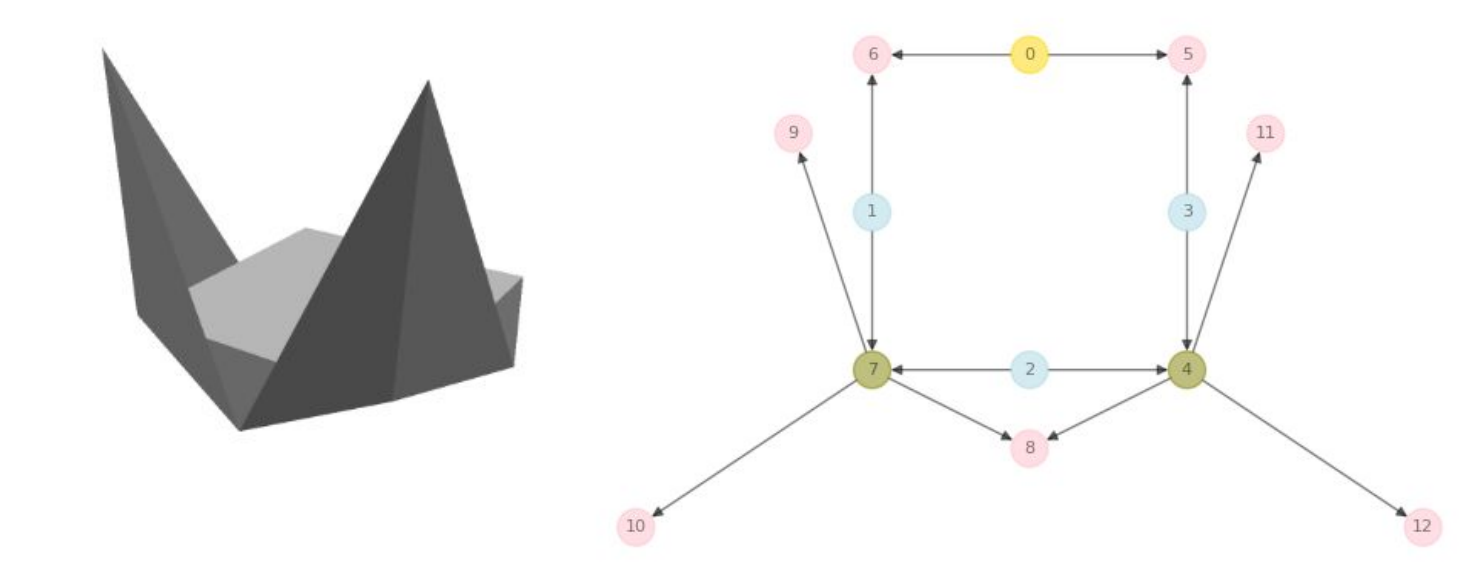}
    \caption{Seeding pattern for the chair fold. Left: the folded shape of the seeding pattern on its own. Right: The graph of the seeding pattern.}
    \label{fig:chair_seed}
\end{figure}

We here detail the starting configurations and symmetries enforced in the different shapes. Specifically, for all abstract rewards, we do not allow for additional source actions apart from the starting pattern. We start each pattern with 4 sources arranged in a square (as we also have in the first three shape approximation experiments), but let the algorithm choose the size of the square. Only for the chair do we fix the size of the starting square as the reward is based on the size of this square (which represents the area on which one sits, see Appendix~\ref{appx:chair_reward}). Moreover, for the bucket and table, we enforce $x$, $y$, and $xy$ symmetry axes. Note however that the $xy$ symmetry here can also be omitted, see Section~\ref{appx:additional_imagined_shapes} for corresponding results.
For the shelf, we enforce symmetry across the $x$ and $y$ axis. Note that the visualization in Figure~\ref{fig:abstract_objective_results} is rotated such that the parallel planes are horizontal (as they would be if one were to use it as a shelf). Also here we can drop one of the symmetry axes to get interesting alternative results, see Section~\ref{appx:additional_imagined_shapes}.
Finally, for the chair, we only have one symmetry axis (the $y$ axis). Here we also provide a seeding pattern as visualized in Figure~\ref{fig:chair_seed}. The intuition here is to flip a downward fold into upward spikes to provide a starting point for the backrest.

For all shapes except the chair, we optimize over the folding angle $\rho_0$. For the chair, we fix $\rho_0$ to the maximal folding angle admissible by the seeding pattern.

\newcommand*\rowlegend[{1}]{\begin{minipage}{0.06\columnwidth}
        \rotatebox{90}{\parbox{2.2cm}{\centering #1}}
        \end{minipage}}

\begin{figure*}[ht]
    \centering
    \hspace*{1cm}
    \rowlegend{Pyramid}
    \hfill
    \begin{minipage}{0.9\linewidth}
        \includegraphics[trim=0 0 0 0, clip,width=0.19\linewidth]{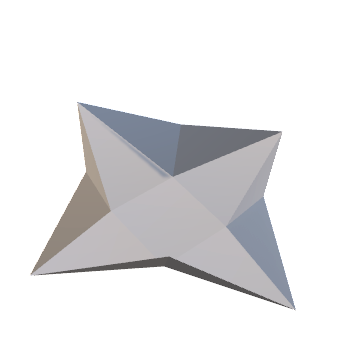}
        \includegraphics[trim=0 0 0 0, clip,width=0.19\linewidth]{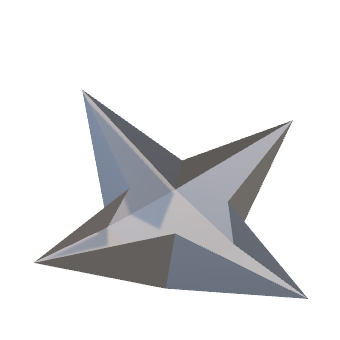}
        \includegraphics[trim=0 0 0 0, clip,width=0.19\linewidth]{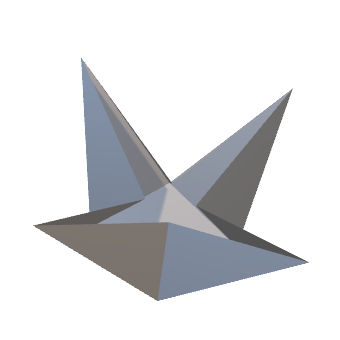}
        \includegraphics[trim=0 0 0 0, clip,width=0.19\linewidth]{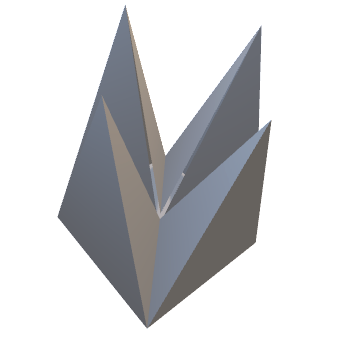}
        \includegraphics[trim=0 0 0 0, clip,width=0.19\linewidth]{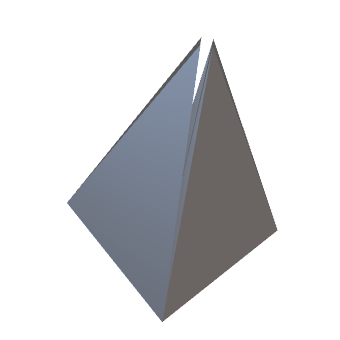}
    \end{minipage}
    \hspace*{1cm}
    \rowlegend{Cube}
    \begin{minipage}{0.9\linewidth}
        \includegraphics[trim=0 0 0 0, clip,width=0.19\linewidth]{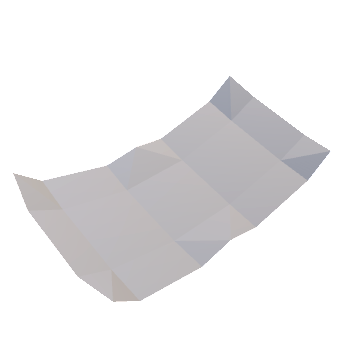}
        \includegraphics[trim=0 0 0 0, clip,width=0.19\linewidth]{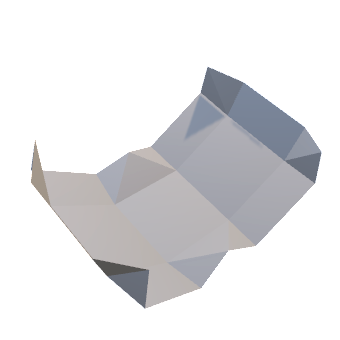}
        \includegraphics[trim=0 0 0 0, clip,width=0.19\linewidth]{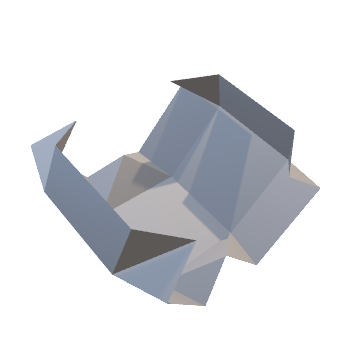}
        \includegraphics[trim=0 0 0 0, clip,width=0.19\linewidth]{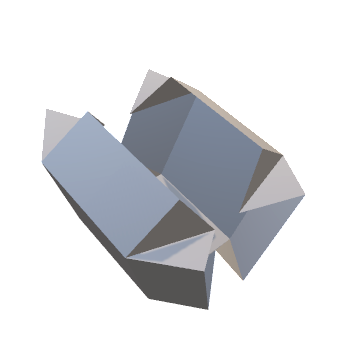}
        \includegraphics[trim=0 0 0 0, clip,width=0.19\linewidth]{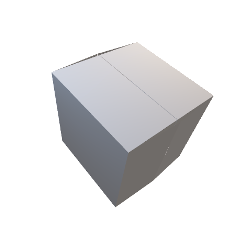}
    \end{minipage}
    \hspace*{1cm}
    \rowlegend{Bowl}
    \hfill
    \begin{minipage}{0.9\linewidth}
        \includegraphics[trim=0 0 0 0, clip,width=0.19\linewidth]{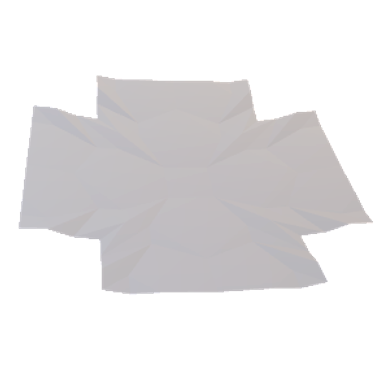}
        \includegraphics[trim=0 0 0 0, clip,width=0.19\linewidth]{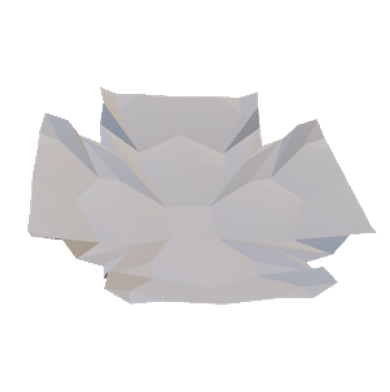}
        \includegraphics[trim=0 0 0 0, clip,width=0.19\linewidth]{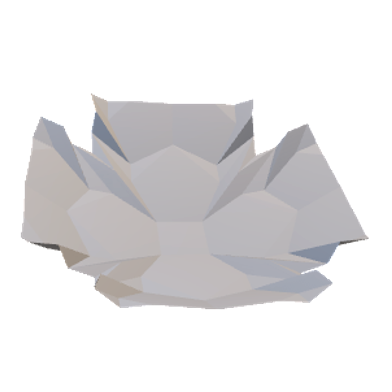}
        \includegraphics[trim=0 0 0 0, clip,width=0.19\linewidth]{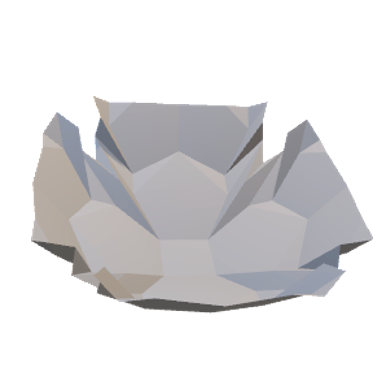}
        \includegraphics[trim=0 0 0 0, clip,width=0.19\linewidth]{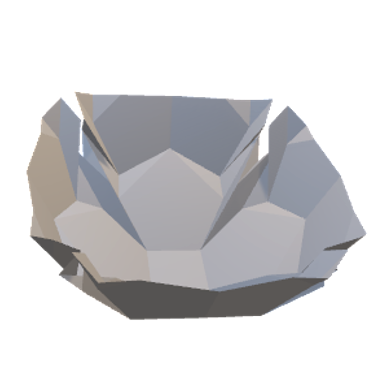}
    \end{minipage}
    \hspace*{1cm}
    \rowlegend{Human Face}
    \begin{minipage}{0.9\linewidth}
        \includegraphics[trim=0 0 0 0, clip,width=0.19\linewidth]{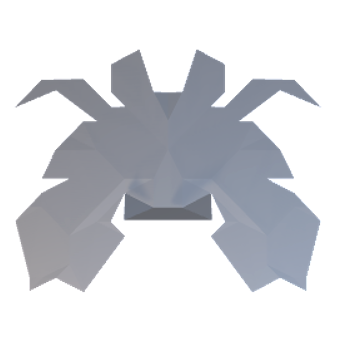}
        \includegraphics[trim=0 0 0 0, clip,width=0.19\linewidth]{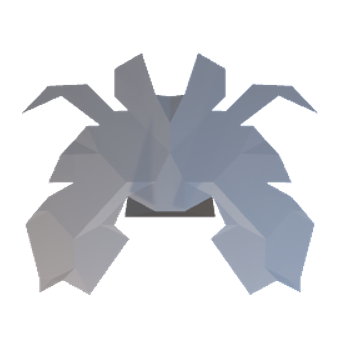}
        \includegraphics[trim=0 0 0 0, clip,width=0.19\linewidth]{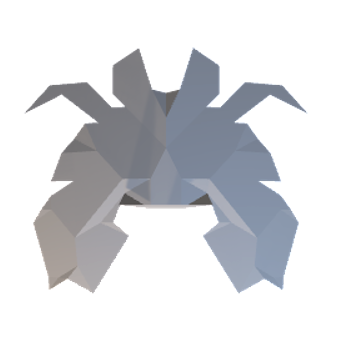}
        \includegraphics[trim=0 0 0 0, clip,width=0.19\linewidth]{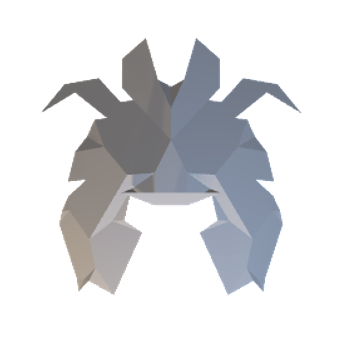}
        \includegraphics[trim=0 0 0 0, clip,width=0.19\linewidth]{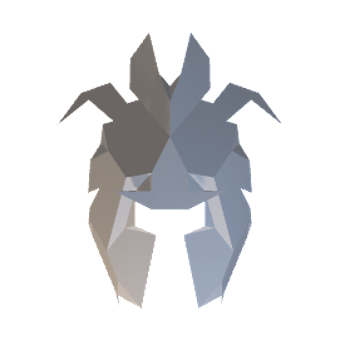}
    \end{minipage}
	\caption{The folding motion of the best approximations for the four targets.}
	\label{fig:folding_motion}
\end{figure*}

\section{Furniture Folding}
\label{appx:furniture_folding}
\begin{figure*}[t]        
\centering
    \subfloat[Bucket crease pattern]{
        \centering
        \includegraphics[width=0.475\textwidth]{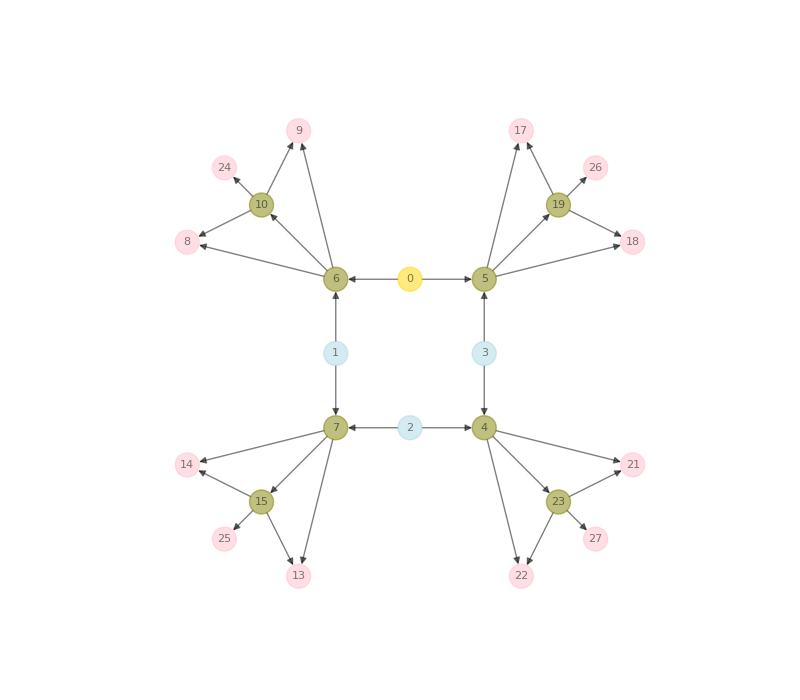}
        \label{fig:imagination_crease_patterns_a}
    }
    \hfill
    \subfloat[Shelf crease pattern]{
        \centering 
        \includegraphics[width=0.475\textwidth]{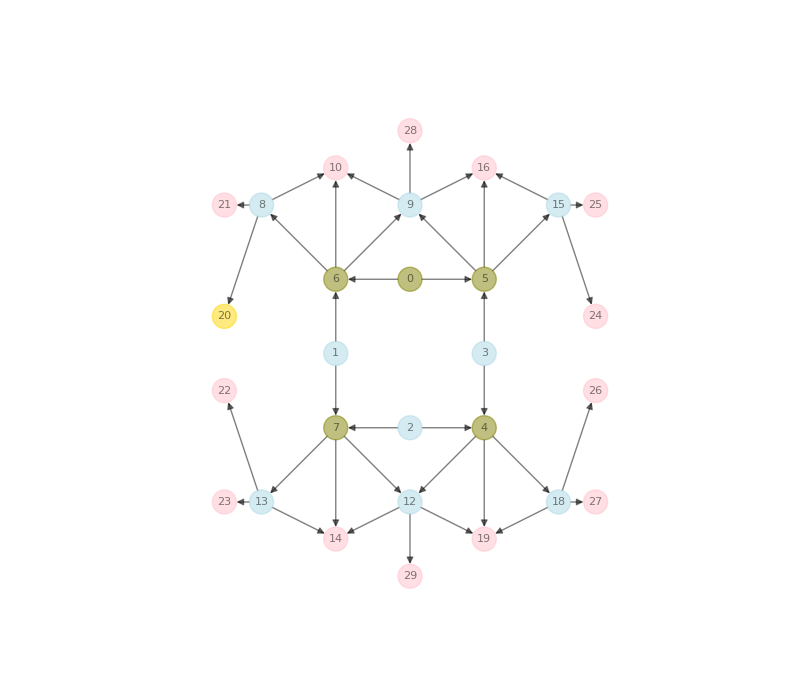}
        \label{fig:imagination_crease_patterns_b}
    }
    \vskip\baselineskip
    \subfloat[Table crease pattern]{  
        \centering 
        \includegraphics[width=0.475\textwidth]{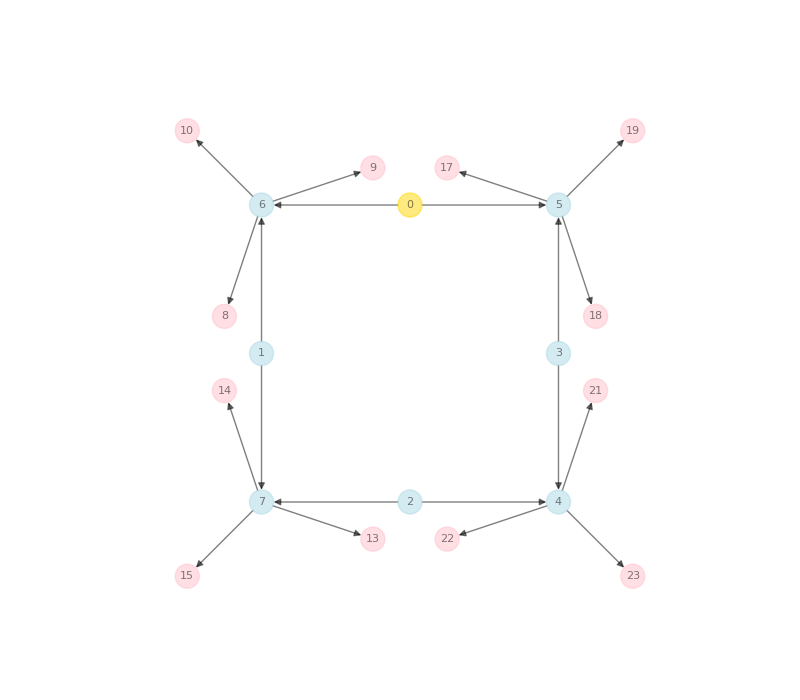}
        \label{fig:imagination_crease_patterns_c}
    }
    \hfill
    \subfloat[Chair crease pattern]{ 
        \centering 
        \includegraphics[width=0.475\textwidth]{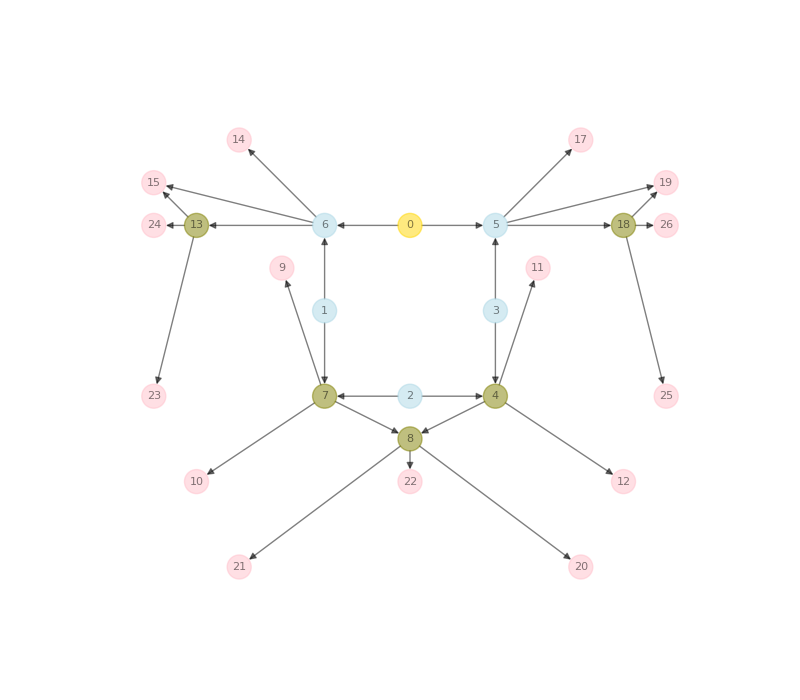}
        \label{fig:imagination_crease_patterns_d}
    }
    \caption{The crease patterns of the imagined shapes: (a) bucket, (b) shelf, (c) table, and (d) chair.} 
    \label{fig:imagination_crease_patterns}
\end{figure*}

\begin{figure*}[t]
    \centering
    \hspace*{1cm}
    \rowlegend{Bucket}
    \begin{minipage}{0.9\linewidth}
        \includegraphics[trim=0 0 0 0, clip,width=0.19\linewidth]{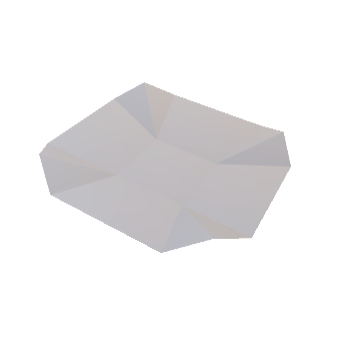}
        \includegraphics[trim=0 0 0 0, clip,width=0.19\linewidth]{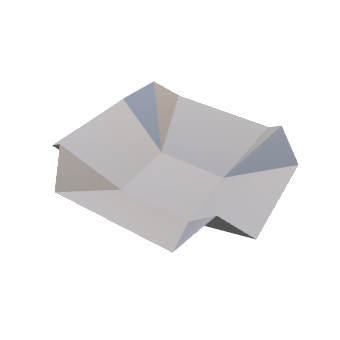}
        \includegraphics[trim=0 0 0 0, clip,width=0.19\linewidth]{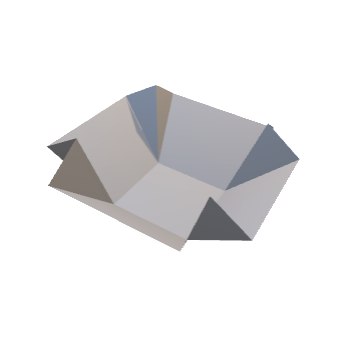}
        \includegraphics[trim=0 0 0 0, clip,width=0.19\linewidth]{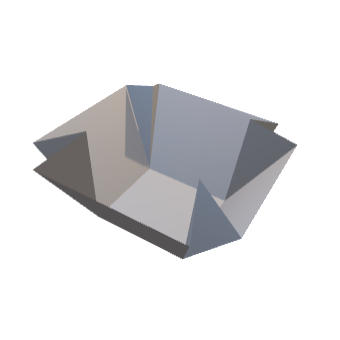}
        \includegraphics[trim=0 0 0 0, clip,width=0.19\linewidth]{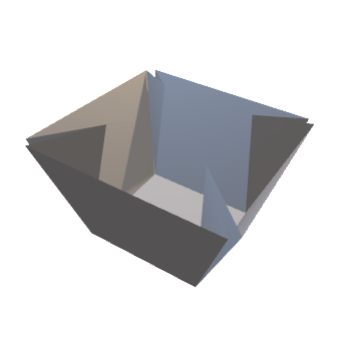}
    \end{minipage}
    \hspace*{1cm}
    \rowlegend{Shelf}
    \hfill
    \begin{minipage}{0.9\linewidth}
        \includegraphics[trim=0 0 0 0, clip,width=0.19\linewidth]{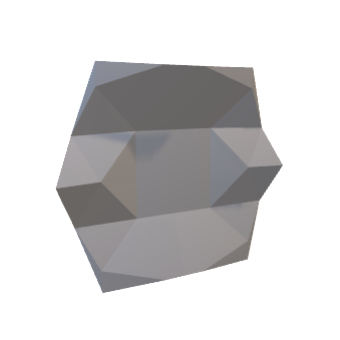}
        \includegraphics[trim=0 0 0 0, clip,width=0.19\linewidth]{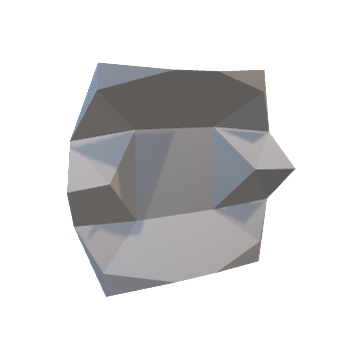}
        \includegraphics[trim=0 0 0 0, clip,width=0.19\linewidth]{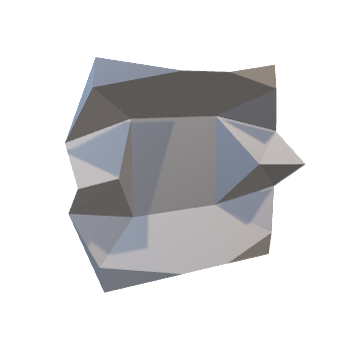}
        \includegraphics[trim=0 0 0 0, clip,width=0.19\linewidth]{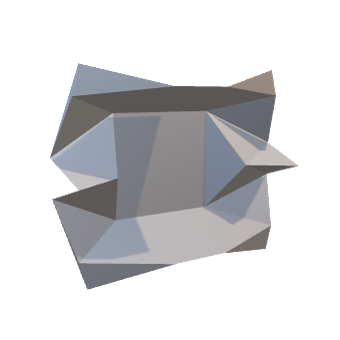}
        \includegraphics[trim=0 0 0 0, clip,width=0.19\linewidth]{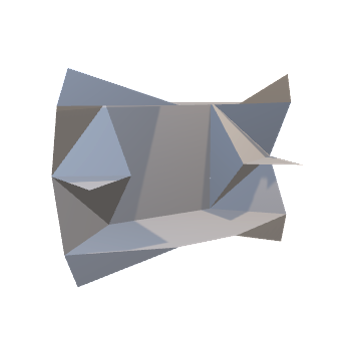}
    \end{minipage}
    \hspace*{1cm}
    \rowlegend{Table}
    \hfill
    \begin{minipage}{0.9\linewidth}
        \includegraphics[trim=0 0 0 0, clip,width=0.19\linewidth]{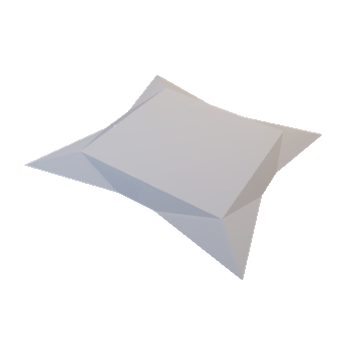}
        \includegraphics[trim=0 0 0 0, clip,width=0.19\linewidth]{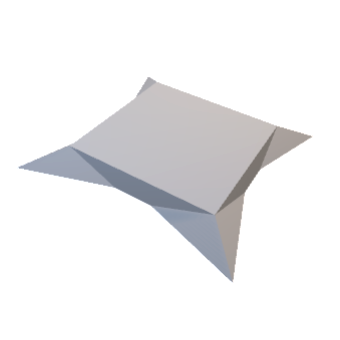}
        \includegraphics[trim=0 0 0 0, clip,width=0.19\linewidth]{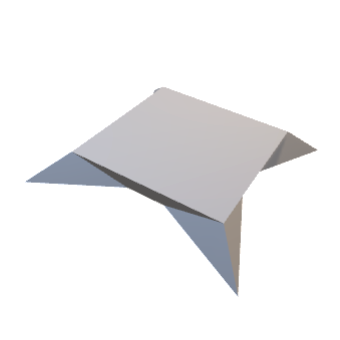}
        \includegraphics[trim=0 0 0 0, clip,width=0.19\linewidth]{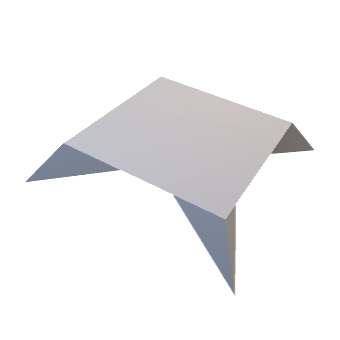}
        \includegraphics[trim=0 0 0 0, clip,width=0.19\linewidth]{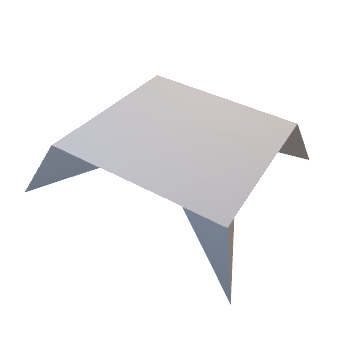}
    \end{minipage}
    \hspace*{1cm}
    \rowlegend{Chair}
    \hfill
    \begin{minipage}{0.90\linewidth}
        \includegraphics[trim=0 0 0 0, clip,width=0.19\linewidth]{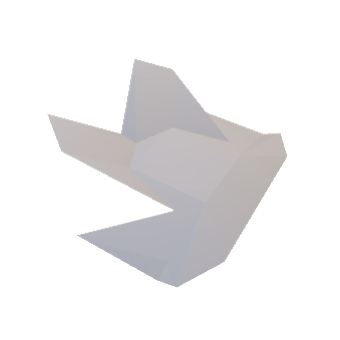}
        \includegraphics[trim=0 0 0 0, clip,width=0.19\linewidth]{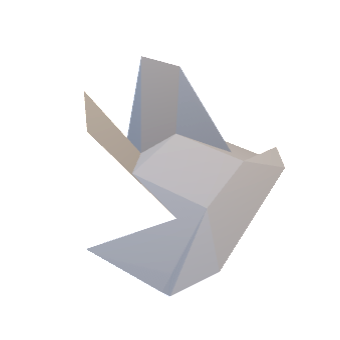}
        \includegraphics[trim=0 0 0 0, clip,width=0.19\linewidth]{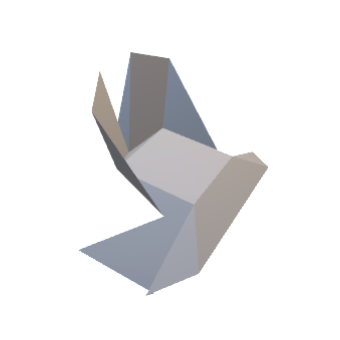}
        \includegraphics[trim=0 0 0 0, clip,width=0.19\linewidth]{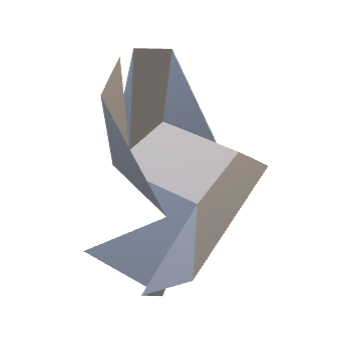}
        \includegraphics[trim=0 0 0 0, clip,width=0.19\linewidth]{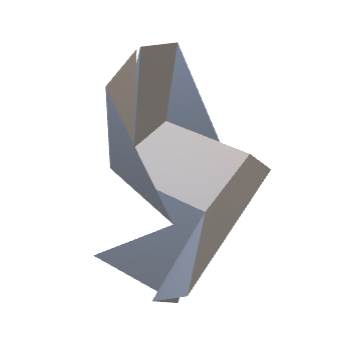}
    \end{minipage}
	\caption{The folding motion of the imagined shapes: bucket, shelf, table, and chair.}
	\label{fig:folding_motion_imaginations}
\end{figure*}

Figure~\ref{fig:imagination_crease_patterns} shows the crease patterns of the everyday objects designed by optimizing our abstract objectives. Meanwhile, Figure~\ref{fig:folding_motion_imaginations} provides a visualization of the folding motion of those crease patterns.

\end{document}